# Deep learning optimal molecular scintillators for dark matter direct detection

Cameron Cook,[1,*] Carlos Blanco,[2,3,†] and Juri Smirnov[1,‡]

[1]*Department of Mathematical Sciences, University of Liverpool, Liverpool, L69 7ZL, United Kingdom*
[2]*Department of Physics, Princeton University, Princeton, New Jersey 08544, USA*
[3]*Stockholm University and The Oskar Klein Centre for Cosmoparticle Physics, Alba Nova, 10691 Stockholm, Sweden*

Direct searches for sub-GeV dark matter are limited by the intrinsic quantum properties of the target material. In this proof-of-concept study, we argue that this problem is particularly well suited for machine learning. We demonstrate that a simple neural architecture consisting of a variational autoencoder and a multilayer perceptron can efficiently generate unique molecules with desired properties. In specific, the energy threshold and signal (quantum) efficiency determine the minimum mass and cross section to which a detector can be sensitive. Organic molecules present a particularly interesting class of materials with intrinsically anisotropic electronic responses and $\mathcal{O}$(few) eV excitation energies. However, the space of possible organic compounds is intractably large, which makes traditional database screening challenging. We adopt excitation energies and proxy transition matrix elements as target properties learned by our network. Our model is able to generate molecules that are not in even the most expansive quantum chemistry databases and predict their relevant properties for high-throughput and efficient screening. Following a massive generation of novel molecules, we use clustering analysis to identify some of the most promising molecular structures that optimize the desired molecular properties for dark matter detection.

## I. INTRODUCTION

In order to look for sub-GeV dark matter, radically new materials must be used for the direct detection of these light particles. While dark matter with weak-scale masses may impart keV of energy into an atomic target—enough to ionize xenon—MeV-scale dark matter can only drive electronic transitions on the order of an eV. Therefore, if we want to probe down to this scale of masses, it is natural to look for materials with transition energies that are at the eV scale, i.e., optoelectronic transitions. Naturally, semiconductors and molecular crystals have emerged as promising materials to use as detector targets. Indeed, results using these materials in the last decade have begun to probe sub-GeV dark matter at ever-smaller cross sections toward benchmark values predicted by cosmologically motivated dark-matter models see for example Refs. [1–3]. However, the specific materials chosen for many of these searches have not been optimized for dark-matter searches but rather are picked due to their practicality of deployment and their relative availability (e.g., EJ-301 scintillator in a pilot direct detection experiment [4]).

Existing experiments and proposals have successfully proven the concept that molecular crystals and semiconductors are well-suited to absorb the energy and momentum delivered in a sub-GeV dark-matter scattering event. However, the target space of viable molecules and crystal structures is enormous and has not been systematically searched. Here, we consider the massive space of organic molecules and attempt to optimize a pair of quantum observables as proxies for the electronic response due to dark-matter-induced electronic recoil, particularly by minimizing transition energies and maximizing oscillator strengths (vertical transition matrix elements).

As shown in Fig. 1, we adopt a generative machine learning (ML) architecture that learns to map the discrete space of string-representations of molecules into a continuous $N$-dimensional vector space known as the latent space. The continuity of the latent space allows for much easier optimization as the network learns to map the structural information of molecules to their location in the latent space. By sampling the latent space and reconstructing vectors within the latent space back into a molecular representation, we can generate new molecules. Our property prediction network can then be used to screen the generated molecules without the need for computationally expensive density functional theory (DFT) calculations. This allows us to use our generative network to find molecules whose properties are optimized for the direct detection of dark matter.

*Contact author: sgccook2@liverpool.ac.uk
†Contact author: carlosblanco2718@princeton.edu
‡Contact author: juri.smirnov@liverpool.ac.uk

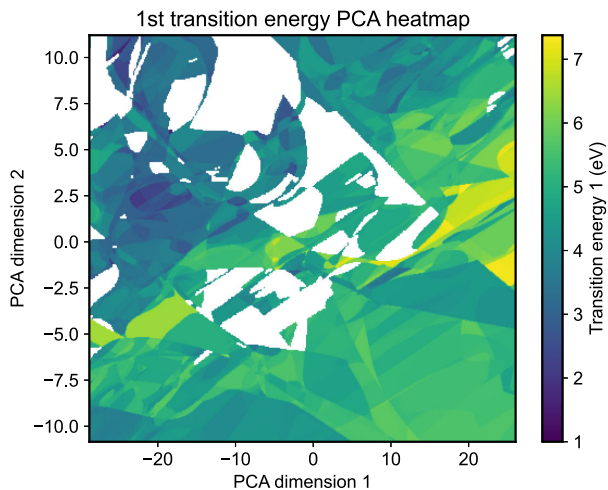

FIG. 1. Visualization of the learned molecular landscape in the variational autoencoder's internal space. Each point corresponds to a molecule, positioned according to the model's most important latent directions. Colors show predicted transition energies, with cooler regions indicating low-energy candidates of interest. Blank regions mark areas where the model did not generate chemically valid molecules. This map illustrates how the generative model organizes chemical space, highlighting promising regions for discovery.

Our search for ideal organic molecules makes use of two ML models and a dataset of organic molecules with precomputed molecular properties, namely the PubChemQC3M dataset. We select 3.3 million from the entire currently available PubChemQC86M dataset [5]. PubChemQC3M, the primary dataset used within this paper is available at [6], more details are available within and at Sec. XI. Our two models, the variational autoencoder (VAE) and multilayered perceptron (MLP) are used in combination such that we can generate new molecules (using the VAE) and are able to predict their corresponding properties that we aim to optimize.

Our generative model is a variational autoencoder (VAE) composed of an encoder and decoder, both implemented as recurrent neural networks (RNNs). Trained on the PubChemQC3M dataset, the VAE learns a continuous latent space representation of molecular structures from their string-based encodings. By sampling and decoding points from this space, we can generate novel molecules. A separate multilayer perceptron (MLP) predicts key quantum properties for these candidates, allowing us to identify those that meet our optimization criteria.

In previous studies, as discussed in Ref. [7], data-mining and machine learning approaches have been successfully applied to property predictions of thermoelectric materials [8], Dirac materials [9,10], topological insulators [11], as well as superconductors [12,13]. In this work we explore the vast space of organic molecules, and identify promising candidates.

In Sec. II we give an executive summary of our strategy and main findings. In Sec. III we give an introduction to the Chemistry required to understand this paper, in Sec. IV we discuss the general methods of sub-GeV direct detection. In Sec. V, we describe our pipeline and the specific network architectures. In Sec. VII, we summarize and discuss our results. In Sec. VI A and VI C, we analyse key molecular structures.

## II. EXECUTIVE SUMMARY

Our main goal is to find molecules that have low ($\sim$eV-scale) transition energies with large [$\sim \mathcal{O}(1)$] corresponding oscillator strengths, which are used as a proxy for the quantum efficiency of the transition following a particle-scattering event. In order to overcome the challenge of sifting through the vast molecular space, we develop a machine learning framework that both generates new molecular candidates and predicts their relevant properties.

Our approach combines two key models:

(1) *A variational autoencoder (VAE)*—a generative neural network that maps the discontinuous space of molecules into a continuous space of probability distributions, enabling the creation of new molecular structures with desired properties.

(2) *A multilayer perceptron (MLP)*—a predictive model that estimates key quantum properties, including transition energy (which determines sensitivity to low-mass dark matter) and oscillator strength (which captures the probability of electronic excitations).

In Fig. 2 we show a schematic of our overall architecture. The novelty of our approach lies in combining a generative model with a regressive model that predicts properties of the newly generated outputs. The VAE architecture learns to encode and decode structures and, in doing so, creates a latent space that allows the creation of new structures. We use a trick to maximize the information content for property prediction by converting the output structures into molecular fingerprints [14,15]. These are bit vectors that tabulate concrete structural and topological information of the molecule. Combining the fingerprints with the latent vectors allows efficient property prediction.

By training these models on a large quantum chemistry dataset (PubChemQC3M), our model can accurately predict energies and oscillator strengths for a wide variety of molecules. We then generate millions of novel molecules and filter them by their transition energy and oscillator strengths. We define threshold values for our region of interest (ROI). Novel molecules in our ROI are then clustered by molecular similarity in order to identify structural backbones/motifs that are associated with enhanced DM detection sensitivity. Our findings indicate that planar polycyclic molecules with asymmetric electron density distributions are particularly promising candidates.

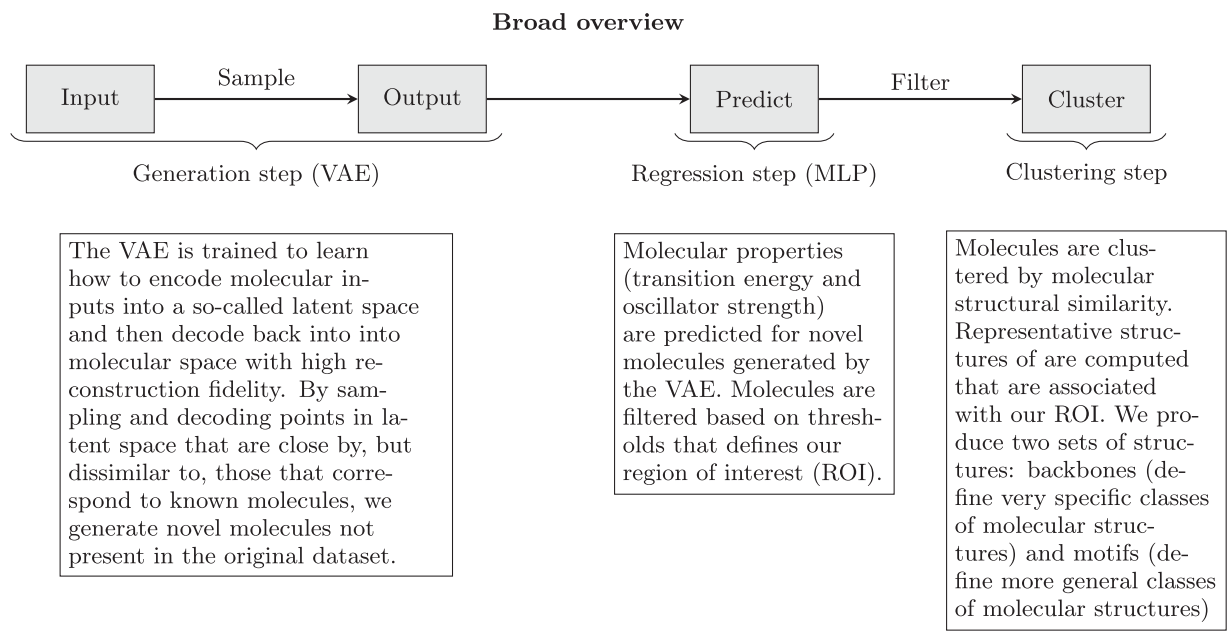

FIG. 2. Schematic overview of the pipeline from the initial generation of novel molecules from known molecular databases to the clustering and generation of key molecular backbones/motifs associated with our ROI.

### A. Key findings

(i) Our machine learning framework successfully generates previously unknown molecular candidates with near-optimal quantum properties for dark matter detection.

(ii) The optimal structures feature delocalized $\pi$-electron systems and asymmetries that enhance electronic transitions.

(iii) A subset of the most promising molecular structures is already known to form crystalline structures, making them viable for practical detector applications.

Our results demonstrate that machine learning is an effective tool for navigating the vast chemical space and identifying optimal materials for dark matter detection. In future work, we will focus on validating our predictions with quantum chemistry calculations and assessing their experimental feasibility.

### III. CHEMISTRY BACKGROUND

In this section, we will provide the basics of organic chemistry needed to understand how carbon-based molecules can play a role in dark matter detection. Organic compounds offer a unique playground for creating materials with optimal electronic properties. This is largely due to the versatility of carbon, which can form different kinds of bonds and a huge variety of structures, from simple chains to more complex systems with delocalized electrons. These properties are particularly important when designing materials that can respond to small energy deposits, such as those expected from interactions with light dark matter particles.

### A. Organic chemistry primer

Organic chemistry is fundamentally the chemistry of carbon-based molecules. What makes carbon, with its four valence electrons, uniquely suited to building an extraordinary diversity of structures is its ability to form four covalent bonds and to adopt different hybridization states depending on its bonding environment.

Figure 3 shows the two most relevant electronic states of carbon in a molecular system. In the simplest case, carbon can be in the $sp^3$-hybridized state [Fig. 3(a)], where its four valence orbitals combine to form four equivalent bonds arranged tetrahedrally forming $\sim 109.5°$ angles. This hybridization underlies the three-dimensional flexibility of molecules like alkanes. On the other hand, the carbon double bond, forms from an $sp^2$-hybridized state [Fig. 3(b)], producing three coplanar bonds at 120° angles. Here, one unhybridized p-orbital participates in $\pi$-bonding, extending above and below the molecular plane formed by the axially bonding $\sigma$-orbitals. The overlap of the extended $\pi$-orbitals inhibit the rotation of the bond making in the molecular structure stiff and planar. Finally, in sp-hybridization, carbon bonds via one tightly binding $\sigma$-bond (forming a

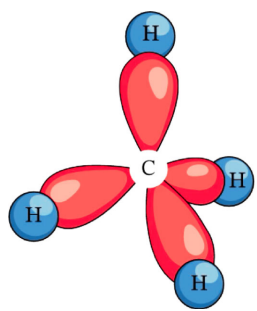

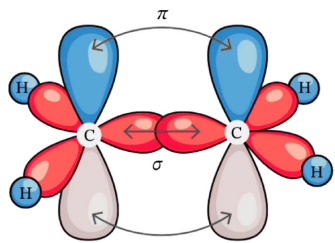

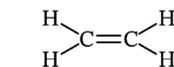

tetrahedral
bonds: 109.5°

planar
bonds: 120°

(a) sp³ hybridisation

(b) sp² hybridisation

FIG. 3.  Illustration of carbon hybridization and $\pi$-conjugation in organic molecules. Top—sp²-hybridized carbon forming saturated, tetrahedral bonds as in alkanes; Bottom—sp²-hybridized carbon enabling planar structures with delocalized $\pi$-electrons above and below the molecular plane, as in benzene; those form the basic building block of extended $\pi$-conjugated systems allowing electron delocalization over multiple carbon atoms, critical for low-energy excitations relevant to dark matter detection.

180° angle) and retains two orthogonal unhybridized p-orbitals leading to a tightly binding linear structure. These different hybridization states allow carbon to be arranged in chains, branched structures, rings, and networks with remarkable mechanical and electronic properties.

A particularly important feature in organic chemistry is the formation of extended $\pi$-electron systems. In molecules with conjugated double bonds—where single and double bonds alternate—the unhybridized p-orbitals from adjacent carbon atoms align and overlap side-by-side, allowing $\pi$-electrons to become essentially delocalized over the network of alternating single and double bonds. Excitations of electrons in these delocalized systems are generally lower energy than those in single bonds. As a result, conjugated organic systems typically exhibit lower excitation energies compared to isolated double bonds or alkanes, and can be optically excited with visible or near-UV photons. This property forms the basis for many applications in organic electronics, photovoltaics, and fluorescent probes.

### B. Structures

Aromaticity is a phenomenon where molecular rings of non-sp³-hybridized states exhibit stabilization to a greater degree than would be expected from conjugation alone (alternating double bonds). Aromatic compounds such as benzene and its derivatives are prime examples and provide a particularly compelling starting point for identifying organic scintillators optimized for dark matter detection. The conjugated $\pi$-electron systems, introduced in the previous section, in benzenelike molecules support low-energy excitations in the few-eV range, enabling sensitivity to dark matter as light as a few MeV. These excitations arise from delocalized electrons occupying molecular orbitals formed by the overlap of $2p_z$ atomic orbitals from, e.g., carbon atoms arranged in a planar hexagonal ring. In such systems, the transitions responsible for scintillation occur within the $\pi$-conjugated molecular orbitals. The relative simplicity and symmetry of benzene's electronic structure, alongside its ubiquity and chemical stability, make it an ideal minimal model for constructing and analyzing molecular form factors relevant for dark matter detection.

Figure 4 shows a benzene ring, an important structure for organic low-mass dark matter detectors, as we will demonstrate in this work. Moreover, the analytic tractability of benzene's wave functions—especially within the linear combination of atomic orbitals (LCAO) approach—allows one to explicitly compute dark matter–electron scattering rates and associated form factors, see Ref. [4]. The six carbon atoms in benzene yield a closed-shell configuration of $\pi$-electrons, whose excitation energies and spatial distributions are well understood. These properties have already enabled the experimental determination of competitive direct-detection limits in a proof-of-principle experiment using benzene derivatives (para-xylene) in low-threshold scintillator cells [4]. Importantly, the theoretical techniques developed for benzene generalize to more complex polycyclic aromatic molecules, allowing its structure to serve as a modular foundation upon which lower-threshold and higher-efficiency targets may be systematically built. This motivates the use of benzene (and para-xylene) as benchmarks in the search for optimal organic targets.

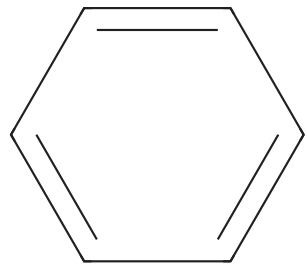

FIG. 4.  A common molecular structure found within a wide array of molecules: a benzene ring with alternating double bonds and a fully delocalized $\pi$−electron system.

### C. Cheminformatics

Machine learning models have proven extremely powerful in processing information represented by languages. In this work, we use symbolic strings with intrinsic syntax rules to represent organic molecules and will demonstrate that our ML architecture creates data representations that allow property prediction and optimization.

Our pipeline uses a character-string molecular representation known as SELFIES (self-referencing embedded strings) [16]. SELFIES is a molecular representation that improves upon the widely used SMILES (simplified molecular input line entry system)—also a character-string molecular representation [17]. Both SELFIES and SMILES encode the structural, elemental, and bond-level information of a molecule as a string of characters. These representations are defined by a collection of syntax rules, making them more like a language as opposed to a featurized numerical representation of molecules.

Since SMILES strings condense complex graphs of elements into information-dense strings, they are widely used in computational chemistry. In particular, they are used in machine learning applications due to their easy conversion to one-hot encodings, which are a common input format in neural networks. One-hot encodings provide an unbiased representation of strings, where each character in the string is represented by a binary vector. These encodings are structured as 2D arrays: one axis represents the alphabet of possible string characters, and the other indicates the position of each character in the string.

However, for generative tasks—i.e., tasks where we aim to produce new molecules outside the original dataset— SMILES falls short. As demonstrated in Ref. [16], the SMILES representation is not robust under point mutations: replacing a single character in the string with another from the alphabet often yields an invalid molecule. This is problematic because generative processes operating on such strings naturally introduce mutations of this kind. Here, "invalid" refers to chemically impossible molecules that violate valence rules (e.g., an element having more bonds than allowed). In contrast, every SELFIES string is

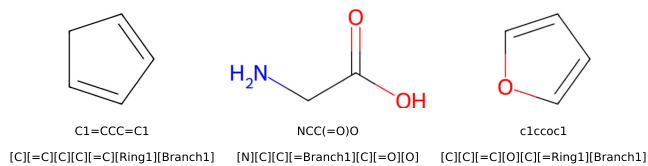

C1=CCC=C1                    NCC(=O)O                c1ccoc1
[C][=C][C][C][=C][Ring1][Branch1]  [N][C][C][=Branch1][C][=O][O]  [C][C][C][O][C][=Ring1][Branch1]

FIG. 5.   For three molecules we show the comparison between the SMILES (top) and SELFIES (bottom) encoding. Note that the SELFIES encoding has a richer structure that ensures validity of the bond valences.

guaranteed to represent a syntactically valid molecule, as its syntax and vocabulary are defined to enforce chemical validity. This property makes SELFIES particularly well suited to our pipeline, since we aim to generate new, valid (and potentially stable) molecules. Consequently, all SMILES-based databases used in this work are first converted to SELFIES before model training.

Figure 5 shows a comparison between the SMILES and SELFIES. Overall it is key to note that SELFIES is combinatorially complete; i.e., the swapping of any piece of syntax within the string will result in a completely valid molecule. Whereas, this is not seen within the SMILES.

### D. Datasets

In this work we use three key molecular datasets to train, validate, and evaluate our networks.

(i) The QM9 molecular dataset [18] is a comprehensive collection of quantum chemical properties for 133,885 small organic molecules composed of carbon (C), hydrogen (H), oxygen (O), nitrogen (N), and fluorine (F) atoms, with up to nine heavy atoms (nonhydrogen). The dataset is complete in terms of possible, valid, nine-atom arrangements, and provides valuable insights into molecular structures and properties, including atomization energy, ground-state dipole moments, electronic spatial extent, polarisability, HOMO-LUMO gap, and vibrational frequencies. The properties in QM9 are computed using density functional theory (DFT) at the B3LYP/6-31G(2df,p) level of theory, a hybrid functional method that balances computational efficiency with accuracy. Each molecule's geometry is optimized prior to calculating these properties, ensuring consistency and reliability across the dataset. QM9 has become a standard benchmark for evaluating machine learning models in quantum chemistry and materials science, serving as a foundational resource for predictive tasks and theoretical analysis.

(ii) A subset of the PubChemQC86M dataset [5]. The PubChemQC dataset is a large-scale quantum chemistry resource derived from the PubChem database, that contains millions of small organic molecules with diverse structures. This dataset provides extensive quantum chemical properties, including molecular energies, oscillator strengths, electron densities, and vibrational frequencies, calculated at various levels of theory. A significant portion of the dataset utilizes DFT calculations with hybrid functionals, such as B3LYP/6-31G(d). In these cases, the molecular geometries are optimized before DFT property calculations. We adopt the PubChemQC dataset as our main dataset for model training due to its size and diversity. As detailed in Sec. XI, we use 3.3 million molecules out of the 86 million available in PubChemQC6M.

(iii) The CSD_EES_DB (Cambridge Structural Database—Experimental Electronic Structure Database) [19] is a specialized dataset derived from the Cambridge Structural Database, focusing on experimental and electronic structure data for crystalline materials. This database provides detailed crystallographic information, including atomic positions, bond lengths, angles, and unit cell parameters, alongside experimental electronic properties such as band structures, density of states, and charge densities. The data is primarily gathered from high-quality x-ray and neutron diffraction experiments, ensuring accuracy and precision in the structural information. Additionally, electronic structure properties are often supplemented by theoretical calculations, such as DFT to complement experimental data. The CSD_EES_DB database allows us to cross-reference novel molecules in our ROI to find those that could form molecular crystals. In forthcoming work, we will include the ability to form crystals as an additional property for our model to predict.

## IV. SUB-GEV DARK MATTER DIRECT DETECTION

Recent studies have shown that organic molecules are particularly well suited to look for MeV-scale dark matter due in part to the kinematic matching between the characteristic momentum of the electrons in the $2p_z$ orbital and the mean momentum imparted during a recoil, $q \approx (m_\chi/\text{MeV})(v/10^{-6})\mathcal{O}(\text{keV})$ [4,20]. In general, the detection strategy proposed in these studies is to measure the excess photoluminescence emitted by a cold volume of molecules that can be attributed to dark matter-induced molecular excitations. Dark matter scattering with electrons in the molecular ground state $|\Psi_g\rangle$ can produce an electronically excited molecular orbital $|\Psi_f\rangle$ if it can impart the necessary transition energy $\Delta E_f$. If this final state is a singlet excited state $|\Psi_{s_f}\rangle$, then the molecule may return to the ground state through radiative deexcitation, emitting a photon and leading to a detection event.

The probability of dark matter-induced excitation is given by the square of the matrix element called the *molecular form factor*,

$$f_{g \to s_f}(\vec{q}) = \langle \Psi_{s_f}(\vec{r}_1, \ldots, \vec{r}_{n_e}) | \sum_i^{n_e} e^{i\vec{q} \cdot \vec{r}_i} | \Psi_g(\vec{r}_1, \ldots, \vec{r}_{n_e}) \rangle, \quad (1)$$

where the sum is over the $n_e$ electrons in the many-body molecular orbital $\Psi$. We follow a common molecular orbital model where the form of the antisymmetric many-electron wave functions is given by Slater determinants of single-particle states $|\phi_i\rangle$. These $|\phi_i\rangle$ states are the ones that diagonalize the core Hamiltonian neglecting electron repulsion, and which are labeled in order of increasing energy. A typical closed-shell ground state $|\Psi_g\rangle$ is given by the following combination of orbitals,

$$\Psi_G = |\psi_1 \overline{\psi}_1 \cdot \ldots \cdot \psi_{n_e/2} \overline{\psi}_{n_e/2}|, \quad (2)$$

where $|\cdots|$ denotes the antisymmetrized product of the single-particle states and $\tilde{\psi}$ is the opposite spin state as $\psi$. A one-electron singlet excited configuration, where an electron initially in the occupied $\psi_i$ is promoted to the unoccupied $\psi_j$, is given by,

$$\Psi_i^j = \frac{1}{\sqrt{2}}(|\psi_1 \tilde{\psi}_1 \ldots \psi_i \tilde{\psi}_j \ldots \psi_{n_e/2} \overline{\psi}_{n_e/2}| - \quad (3)$$

$$|\psi_1 \tilde{\psi}_1 \ldots \psi_j \tilde{\psi}_i \ldots \psi_{n_e/2} \overline{\psi}_{n_e/2}|). \quad (4)$$

The electron repulsion term in the molecular Hamiltonian will mix these configurations, and the excited energy eigenstates are linear combinations of these $\Psi_i^j$ configurations and can be computed iteratively in, e.g., self-consistent Hartree-Fock methods. In any case, writing the states in this way allows us to see that the molecular form factor can be written as a sum of single-particle interactions,

$$\begin{aligned} f_{g \to s_f} &= \sum_{ij} d_{ij}^{(n)} \langle \Psi_i^j | e^{i\vec{q} \cdot \vec{r}} | \Psi_G \rangle \\ &= \sqrt{2} \sum_{ij} d_{ij}^{(n)} \langle \psi_j(\vec{r}) | e^{i\vec{q} \cdot \vec{r}} | \psi_i(\vec{r}) \rangle, \end{aligned} \quad (5)$$

where $d_{ij}$ is the coefficient of the $\Psi_i^j$ configuration in the expansion of the $\Psi_{s_f}$ singlet excited state. In general, these matrix elements can be very difficult to compute since 1. the expansion of $\Psi$ can be very large, and 2. the integrals are over rapidly oscillating functions.

Now that the molecular form factor is written as a sum of single-particle scattering matrix elements, we can move on to computing the excitation rate in a material due to the dark matter flux expected in the lab. The differential scattering cross section can be factored as follows,

$$\frac{d\sigma_e}{dq^2} = \frac{\bar{\sigma}_e}{4\mu_{\chi e}^2 v^2}|F_{DM}(\vec{q})|^2, \qquad (6)$$

where $\mu_{\chi e}$ is the reduced mass between the DM, with mass $m_\chi$, and the electron with mass $m_e$. The relative lab-frame velocity is $\vec{v}$, and $|F_{DM}|^2$ is the so-called dark matter form factor that parametrizes the momentum dependence of the SM-DM matrix element.

We define the fiducial cross section $\bar{\sigma}_e \equiv \frac{\mu_{\chi e}^2}{16\pi m_\chi^2 m_e^2}\langle|\mathcal{M}(q_0)|^2\rangle$, where the free-particle spin-averaged scattering matrix element $\mathcal{M}(q_0)$ for $\chi - e$ is evaluated at the reference momentum $q_0 = \alpha m_e$. In general, a simple spin-independent scattering matrix element is proportional to the propagator of some new force mediator with mass $m_\phi$ such that, $\mathcal{M} \propto 1/(m_\phi^2 + q^2)$. The dark matter form factor is then given by

$$F_{DM}(q) = \frac{\alpha^2 m_e^2 + m_\phi^2}{q^2 + m_\phi^2}. \qquad (7)$$

Note that in the limit of a mediator that is much heavier than the momentum scale $q \approx (m_\chi/\text{MeV})(v/10^{-6})\mathcal{O}(\text{keV})$, the form factor tends to unity $F_{DM}(q) \to 1$ and corresponds to a contact interaction. Conversely, a sufficiently light mediator yields the opposite limit, $F_{DM}(q) \to (\alpha m_e/q)^2$ and corresponds to a long-range interaction.

The observed excitation rate of a detector with $N_T$ electrons in $\Psi_g$ is given by the following,

$$\Gamma = \frac{\Phi_{BF} N_{mol} \rho_\chi}{m_\chi} \frac{\bar{\sigma}_e}{\mu_{\chi e}^2} \sum_{i=1} \int \frac{d^3\vec{q}}{4\pi} \int d^3\vec{v} f_\chi(\vec{v})$$
$$\times \delta\left(\Delta E(s_f) + \frac{q^2}{2m_\chi} - \vec{q}\cdot\vec{v}\right) F_{DM}^2(q)|f_{g\to s_i}(\vec{q})|^2. \qquad (8)$$

Here, $\rho_\chi = 0.4 \text{ GeV}/\text{cm}^3$ is the local mass density of dark matter. The six-dimensional kinematic integral is over the imparted momentum $\vec{q}$, and DM velocity $\vec{v}$, whose local distribution in the lab frame is given by $f_\chi(\vec{v})$, and $\Delta E(s_f)$ is the excitation energy each $\Psi_{s_f}$ above the ground state.

Finally, $\Phi_{BF}$ is the bulk fluorescence quantum yield which quantifies how often one expects that a photon emitted through radiative deexcitation is able to free stream through macroscopic distances in the molecular material. In other words, $\Phi_{BF}$ quantifies how self-transparent the material is to its own fluorescence. In previous work, this quantum yield is adopted from empirical measurements, and in general this can be of order unity for organic crystals. However, $\Phi_{BF}$ is fundamentally dependent on the Stokes shift between the absorption bands and emission bands of the material, i.e., the energy lost to nuclear motion before radiative deexcitation.

The DM velocity distribution in the Galactic rest frame is given in the standard halo model [21],

$$f_{SHM}(\vec{v},t) = \begin{cases} \frac{1}{N}\left(\frac{1}{\pi v_0^2}\right)^{3/2} e^{-v^2/v_0^2}, & v < v_{esc}, \\ 0, & v \geq v_{esc}. \end{cases} \qquad (9)$$

where $N$ is a normalization factor, $v_0 \approx 220 \text{ km/s}$ [22], and $v_{esc} \approx 544 \text{ km/s}$ [23]. The velocity distribution in the lab frame is related to $f_{SHM}(\vec{v},t)$ by the Galilean transformation, $f_{lab}(\vec{v},t) \simeq f_{SHM}(\vec{v} + \vec{v}_\odot(t) + \vec{v}_{earth}(t))$, where $\vec{v}_\odot(t)$ is the velocity of the Sun in the Galactic rest frame and $\vec{v}_{earth}(t)$ is the velocity of the earth in the Solar rest frame.

We note that the above calculation implies the following three desirable properties for a molecule being used as a detector target,

(i) a molecular form factor $|f_{g\to s_i}(\vec{q})|^2$ that is as large as possible, to maximize the rate at a given $\bar{\sigma}_e$.

(ii) a threshold transition energy $\Delta E_{s_f}$ that is as small as possible, in order to minimize the lowest kinematically accessible $m_\chi$.

(iii) $\Phi_{BF}$ as close to 1 as possible to maximize the visibility of the signal.

In this paper, we will focus on the first two since they are governed by the electronic molecular orbitals. The third will be the subject of subsequent work.

Note that we consider the transition dipole moment for the molecular transition of interest as a proxy for the molecular form factor. Expanding the exponential in Eq. (5) we see that the leading order nonvanishing term in the expansion is the orbital overlap integral over $\vec{r}$, which is just the transition dipole moment,

$$\langle\psi_i|e^{i\vec{q}\cdot\vec{r}}|\psi_G\rangle \approx \vec{q}\cdot\langle\psi_j|\vec{r}|\psi_i\rangle + \mathcal{O}(q^2)$$
$$\approx \vec{q}\cdot\langle\vec{r}\rangle_{0,i}, \qquad (10)$$

and its square it proportional to the oscillator strength of the transition. The oscillator strength, shown below, is a dimensionless quantity with no units and heuristically, it represents the probability of a transition occurring.

$$f'_{0,i} = \frac{2m_e\Delta E_{0,i}}{3}|\langle\vec{r}\rangle_{0,i}|^2. \qquad (11)$$

So, we find that the low-momentum behavior of the molecular form factor for dark matter scattering is governed by the oscillator strength of the vertical transition, i.e., the optical absorption oscillator strength,

$$|f_{g\to s_f}|^2 \sim \frac{q^2 f'_{0,i}}{m_e \Delta E_{0,i}}. \qquad (12)$$

Note that the oscillator strength is a dimensionless positive-definite quantity and does not carry a square due to convention.

## V. METHODS

In this section, we discuss the details of our pipeline, particularly the machine learning models and the ways in which they were employed and validated.

In Sec. VA, we describe our specific network architectures. In Sec. VB, we demonstrate our algorithm used to generate our long list of molecules which fit our selection criteria. Finally, in Sec. VC, we set out the clustering algorithms used to find the key molecular structures associated with our region of interest.

### A. Network architecture

Our pipeline requires two machine learning components: a generative and a predictive component. Generative machine learning models aim to learn the distributions of input datasets and then generate new samples from these distributions. Key generative architectures include normalizing flows, generative adversarial networks (GANs), and variational autoencoders (VAEs). Many predictive networks are types of multilayered perceptrons (MLPs) or multilayered fully connected neural networks. Unlike generative models, these networks differ mainly in their layer types. Our options for predictive models included graph neural networks (GNNs), convolutional neural networks (CNNs), recurrent neural networks (RNNs), and standard MLPs. While we use RNNs in our VAE architecture, they are best-suited for variable-length data. By using molecular fingerprints, which are bit vectors of constant length, we can use the simple input layer of MLPs. GNNs can be thought of as a generalization of CNNs (e.g., from grids to general graphs). Although GNNs achieve high accuracy in molecular property prediction, they are generally more memory-intensive than simple MLPs and, for graphs derived from grammar-based representations (e.g., SMILES or SELFIES), offer improvements insufficient to vindicate their memory intensity. We opt for the simplest solutions, an MLP, and show that it performs accurately enough for our purposes.

On the generative front, we choose to use a VAE. VAEs were preferred since they, unlike normalizing flows and GANs, allow for specific sampling of molecules, i.e., we can control how similar we want our new, sampled molecules to be to inputted ones. Again, we use RNNs as our encoder and decoder networks due to their proficiency in learning semantic rules and ability to deal with variable length vectors.

#### 1. Variational autoencoder

The variational autoencoder (VAE) is composed of two networks, the encoder and the decoder, and in our case, each is an RNN. In Fig. 6, we show the constituent subnetworks of the VAE, and sketch the training process. The VAE's ability to be generative relies on learning a mapping between the typically complex and disjointed data

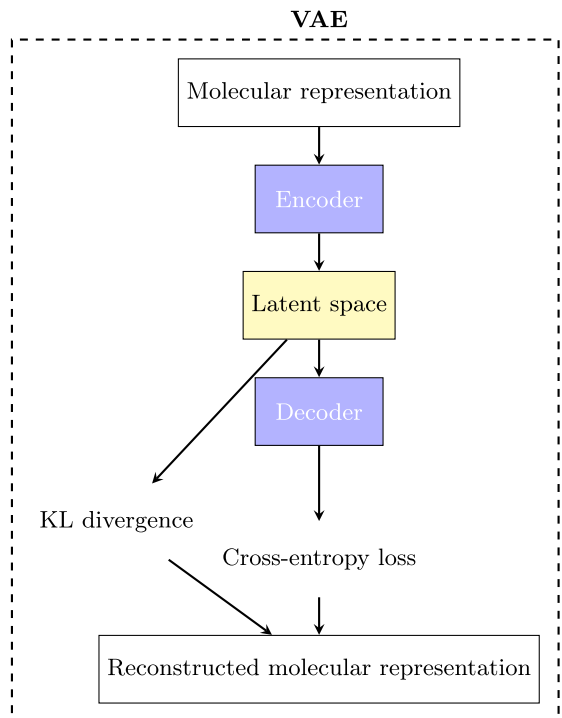

FIG. 6. Network architecture of the variational autoencoder, consisting of an encoder and a decoder network, which are cotrained to create an N-dimensional latent space. The encoded structures can be reconstructed based on a probabilistic process. The network can generate genuinely new structures well outside its training sets.

distribution (in x-space) and a simpler latent prior distribution, such as a Gaussian (in z-space).

For our purposes, objects in x-space are the SELFIES representations of molecules. The z-space, which is a n-dimensional vector space called the latent, hosts two kinds of objects: 1. the outputs of the encoder and 2. the inputs of the decoder. The output of the encoder is a multi-Gaussian distribution $\mathcal{N}(\boldsymbol{\mu}, \boldsymbol{\sigma})$ defined by two N-dimensions vectors, $\mu$ and $\sigma$. Samples $\mathbf{z}$ from these distributions are also n-dimensional vector, and are the inputs of the decoder. From here on, we will refer to $\mathbf{z}$ as latent vectors. We note, that while a given $\mathbf{z}$ may be referred to as the "latent representation" of a molecule $\mathbf{m}$, it should be understood to be a nonunique sample from the encoded distribution (true latent representation) $\mathcal{N}(\boldsymbol{\mu}, \boldsymbol{\sigma})$. By sampling from this distribution to generate new latent vectors $\mathbf{z}$, the VAE can produce similar yet novel molecules. After training the VAE, orthogonal directions in the latent space capture abstract information about the input data. Since our model is trained on SELFIES, the latent vectors should correspond to encoded SELFIES strings. In other words a given SELFIES should be described by one $\mu$ and one $\sigma$ and, by sampling from the distribution defined by these vectors, we can generate new but similar latent vectors $\mathbf{z}$ which would represent new but similar SELFIES.

More explicitly, the encoder maps x-space to latent space, and the decoder maps back. And so, by sampling latent vectors **z** allows decoding into similar but new SELFIES. Throughout this paper, we reference three key vectors:

(i) $\mu$: This is the vector that best represents an input encoding (the mean of the distribution). In our case, one can imagine it as the best mapping of a one-hot encoding to the latent space. If given to the VAE decoder, this molecule has the highest likelihood of returning the original input. Every $\mu$ is an $N$-dimensional vector where $N$ corresponds to the number of dimensions in the latent space.

(ii) $\sigma$: This is the variance vector of a given $\mu$. Each $\sigma$ corresponds to the standard deviation of its corresponding diagonal multi-Gaussian with mean $\mu$. Heuristically, the encoder attributes to every molecule a region in latent space. This region is centered at a vector $\mu$, whose elements encode molecular features, while $\sigma$ determines the size of the area in latent space where similar such features can be found.

(iii) **z**: This is the vector sampled from the distribution defined by $\mu$ and $\sigma$. Sampled vectors ought to correspond to molecules which are new but similar to the input molecules, corresponding to $\mu$.

### 2. Multilayered perceptron

In Fig. 7, we present the second network used in our approach: the multilayer perceptron (MLP). The key feature of the MLP is its input. First, we use the trained VAE to generate a representative $\mu$ from an input molecule (SELFIES). Additionally, we convert the input molecules (SMILES) into several types of molecular fingerprints. $\mu$, along with the fingerprint representations, are then fed into the MLP, which predicts the molecular properties of the corresponding molecule. We find that this method allows us to maximize the information content of the molecular input and leads to a severely boosted property reconstruction efficiency of the MLP.

### B. Seeded sampling algorithm

The primary objective of the seeded sampling algorithm is to find molecules with desirable molecular properties by generating a large number of candidates and picking out molecules that fit our criteria. We choose specific molecular 'seeds' in order to increase the number of strong molecular candidates generated. Within the context of this paper, we refer to molecular seeds as molecules used for sampling within the latent space of our VAE.

This method was preferred to a gradient descent method using an MLP trained on the latent space since the gradient descent algorithm is likely to push into ill-defined regions, unknown to the VAE (with low training data density), producing nonsensical results. That is, the algorithm would spot elements of the latent space associated with desirable molecular properties and move well beyond regions near the latent representation of the training data. We were unable to construct a means to limit the gradient descent algorithm to well-defined regions only, as in a probabilistic setup this definition is blurry.

As shown in Fig. 8, the seeded sampling algorithm ultimately consists of generating many molecules and then predicting their molecular properties, collecting ones that fit a set of predefined criteria. Figure 23 in Appendix A further expands upon Fig. 8. Maximizing our output number is easy since one can increase the number of latent vectors sampled per input molecule, this process is only limited by the available computing resource. As such, we employ a variety of tricks to increase the rate of "ideal molecules" (molecules that pass our selection criteria) produced per unit time.

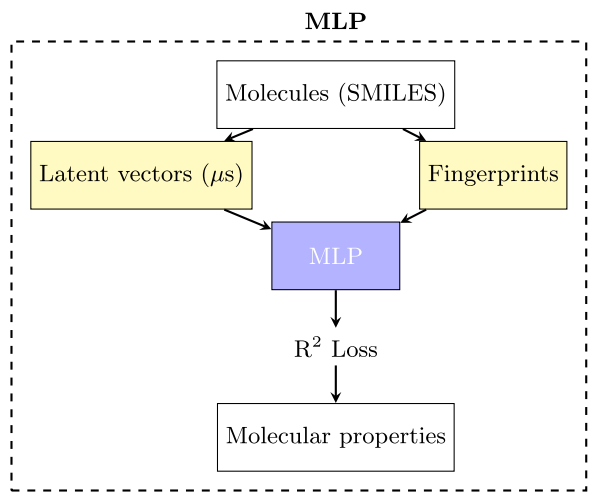

FIG. 7. This diagram shows the MLP trained to predict molecular properties. The MLP shows optimal performance when trained on the $\mu$s outputted by the encoder and molecular fingerprints. Molecular fingerprints are higher information content structures generated from the reconstructed SMILES from the latent space. In our case, there are three types of fingerprints used: Morgan fingerprints, Daylight fingerprints, and mol2vecs. References to the Morgan fingerprints, Daylight fingerprints, and mol2vec can be found at: [14,15] respectively.

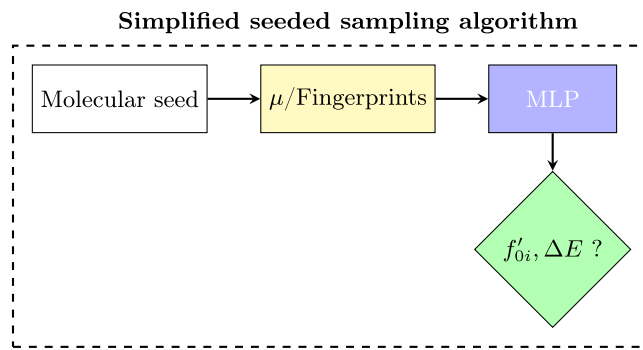

FIG. 8. Simplified diagram to show the seeded sampling process. Many **z**s are sampled around an inputted molecular seed. These **z**s are converted to SMILES where the $\mu$ and fingerprint representations can be fed into the MLP. The algorithm asks which molecules have desirable molecular properties and appends them. A full description of this process can be found in Appendix A.

### 1. Molecular seeds

Molecular seeds refer to the molecules used for sampling within our VAE and, by extension, the seeded sampling algorithm. Since our methodology revolves around using the VAE to generate as many molecules with high oscillator strengths and low transition energies as possible, we have two choices:

(i) Increase yield of seeds
(ii) Decrease samples made per seed

I.e., we can probe each molecule within our dataset and sample just a few molecules per seed or we can focus on very specific seeds and generate many molecules per seed. We focus on the latter approach. And, as such, we develop an algorithm to choose seeds with the highest yield, i.e., highest number of output molecules with desirable properties.

Sampling 1000 latent vector representations ($zs$) per molecular seed, expressed as a $(\mu, \sigma)$, we push all $3.3M$ molecules in the PubChemQC3M dataset through the seeded sampling algorithm. The output of the seeded sampling algorithm will be a set of molecules predicted to have low transition energies and high oscillator strengths. By tagging the output molecules from the seeded sampling algorithm with which molecular seed they were generated from, we can quantify the yield of molecular seeds. That is, we can figure out how many molecules with desirable properties were generated from a given seed. We then classify molecular seeds as being high or low yield. We define a high yield molecular seed as being one which generates at least 3 molecules with desirable properties, i.e., 3 molecules sampled from the given molecular seeds were classified as molecules with desirable properties. Taking all high yield molecular seeds, we are left with a list of 8800 molecules.

Despite the high yield molecular seed list constituting only 0.2% of the PubChemQC3M dataset, it generates almost all of the molecules with desirable molecular properties generated via the seeded sampling algorithm by the entirety of the PubChemQC3M dataset. For our final seeded sampling run, we take the high yield molecular seed list and sample 100,000 latent vectors per seed. Via this method we explore the relevant molecular space by sampling around a small percentage of molecules in our dataset.

### 2. Sampling region

A strong molecular seed list increases the number of molecules with desirable properties outputted by the seeded sampling algorithm, for a given time, by an extraordinary amount. However, if molecules are sampled too closely to the input molecular seed, i.e., the $zs$ are too close to the $\mu$, then we will waste time by generating many nonunique molecules. Therefore, we limit the region around which we sample from the molecular seed. This translates to cutting away at the Gaussian used to generate our distribution of $zs$ such that we only sample within the regions of $[1\sigma < |\mathbf{z} - \mu| < 2\sigma]$.

Figure 9 shows the uniqueness of sampled molecules as a function of the absolute outer radius on the truncated Gaussian distribution used to sample latent vectors ($zs$). This shows that our generative region is ideal, since when sampling from molecular seeds, we produce a large number of unique $zs$ which are not too different from the original input. Furthermore, as mentioned in step 6 in Appendix A, we can remove degenerate reconstructions. This helps us, particularly, by reducing the amount of time constructing the fingerprints, which is the most computationally expensive part of this process.

### C. Clustering algorithms

The seeded sampling algorithm (SS) works effectively to produce a large number of molecules with desirable molecular properties, outputting millions of molecules in our case. However, for our purposes, it is necessary that we narrow this list of molecules to a short list of characteristic molecules that effectively represent the generated data. To this end, we cluster the outputs of the seeded sampling algorithm and then derive representative molecules of the clusters.

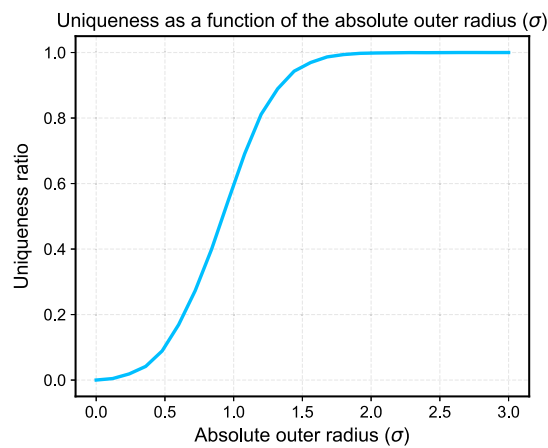

FIG. 9. This plot shows how the mean uniqueness of outputs from the decoder changes as a function of the absolute outer radius of the Gaussian distribution used to sample latent vectors ($zs$). The plot was generated for one molecular seed but reflects the distribution among all other molecular seeds: the further out from the center, the less similar the sampled distribution is to the original input molecule.

We have experimented with a number of clustering and cluster analysis techniques with the goal of accurately grouping the SS outputs. In the end, we have identified two clustering methods that yield complementary results: one provides more general structures, while the other offers more specific ones. Both methods, together, provide a representative description of our dataset by describing key molecular structures associated with our region of search and then providing suggestions for more optimum molecular structures.

`BitBIRCH` clustering is used to identify general molecular structures, while `single-linkage` clustering is employed to find more specific molecular structures. Structures identified by `BitBIRCH` clustering are referred to as molecular `motifs`, whereas structures identified by `Single-linkage` clustering are called molecular `backbones`.

The `BitBIRCH` [24] clustering algorithm is a modification of the balanced iterative reducing and clustering using hierarchies (BIRCH) algorithm [25]. `BitBIRCH` and `BIRCH` both make use of cluster trees to sort molecular inputs by specific properties. Nodes (leaves) on these BIRCH trees correspond to clusters, information about specific clusters is represented as vectorlike objects. A typical vectorlike object contains information like the number of elements, the linear sum, and the squared sum of elements in the cluster. Because of this, `BIRCH` cannot effectively cluster molecule fingerprints, which are expressed as 1s and 0s, since the information within the molecular fingerprints is stored as the pattern of 1s and 0s. I.e., the sum of molecular fingerprints is not meaningful information.

Instead, `BitBIRCH` uses a centroid measure in favor of a squared sum measure, i.e., the mean molecular fingerprint of a cluster. Then, using the vectorlike descriptors of the clusters (the number of elements, the linear sum of elements, and the centroids), `BitBIRCH` defines another measure: the binary clustering features (BCFs). BCFs are vectors of probabilities corresponding to the distribution of molecular features across the cluster, where each component represents the likelihood that a specific molecular feature is present.

New molecular fingerprints are compared to cluster BCFs and if the similarity (commonly Tanimoto) exceeds a specific threshold, the molecule will be accepted into the cluster. If the molecule is accepted to the cluster then the cluster BCFs must be recalculated. If the molecule is not accepted to any cluster then it will be sent to form a new cluster. Note that this clustering process does not require computing a 2-D distance/similarity matrix.

The `Single-linkage` clustering applied here works by taking a similarity matrix (2D array of similarities) and linking every element that has a similarity above a given threshold (0.7 in our case). Clusters are formed from all connected links. For our data, the similarity matrix is based on the Tanimoto similarity between each molecule in the seeded sampling output and every other molecule. Explicitly, the Tanimoto similarity measures the number of shared features between set A and B divided by the total number of features within both set A and B.

In practice, we find all links by using a `Depth-first` search algorithm which begins with a molecule and follows the link chain until there are no links left, thus forming the cluster. Good introductions to `Single-linkage` clustering and `Depth-first` search algorithms can be found at [26,27], respectively.

## VI. STRUCTURE-DRIVEN ANALYSIS

The seeded-sampling algorithm generates an enormous number of molecules which match our criteria (in our ROI). We can use this large-scale dataset of novel molecules in our ROI to determine what molecular properties are related to "optimal" molecular structures. To this end, we use a structure-driven analysis methodology to determine key molecular structures and construct representative molecules strongly associated with them.

In this section, we will discuss the outputs of the two clustering algorithms used. That is, we will discuss the different motifs and backbones (key molecular structures) found from our seeded sampling algorithm. Importantly, in this work the motifs are more general classes of molecular structures that we find to be associated with our ROI. Whereas backbones correspond to specific structures that can be thought of as best performers. Optimal motifs are generated via the BitBirch clustering algorithm and optimal backbones are generated via the single-linkage clustering algorithm. Beyond this, we also use the found optimal motifs to cross-reference a database of known molecular crystals, yielding motifs in our ROI whose representative structures are known crystals.

### A. Motif-finding analysis

Our motif-finding analysis consists of three main stages: clustering, scaffold construction, and cross-section analysis. Overall, our motif driven clustering process is detailed in Fig. 10.

#### 1. Initial clustering

Beginning with a dataset formed from the outputs of the seeded sampling algorithm, we perform a cut on molecules with synthetic accessibility scores (SA) scores [28] greater than 5. The SA score is a composite score of a variety of smaller molecular property scores (e.g., molecular property score) which serves as a proxy for experimental synthetic feasibility. The score ranges from 1–10, with 10 being the most synthetically infeasible and 1 being the most synthetically feasible, as such we assume that the molecules that are left are likely experimentally synthesizable. Using `RDKit`, we calculate the molecular pattern fingerprints for all molecules in the candidate data set. These fingerprints are bit vectors of length 1028. The fingerprints are then used to cluster the molecules according to `BitBIRCH` with a branching factor of 0.5 and a Tanimoto similarity threshold of 0.75. These parameters balance cluster granularity with computational efficiency.

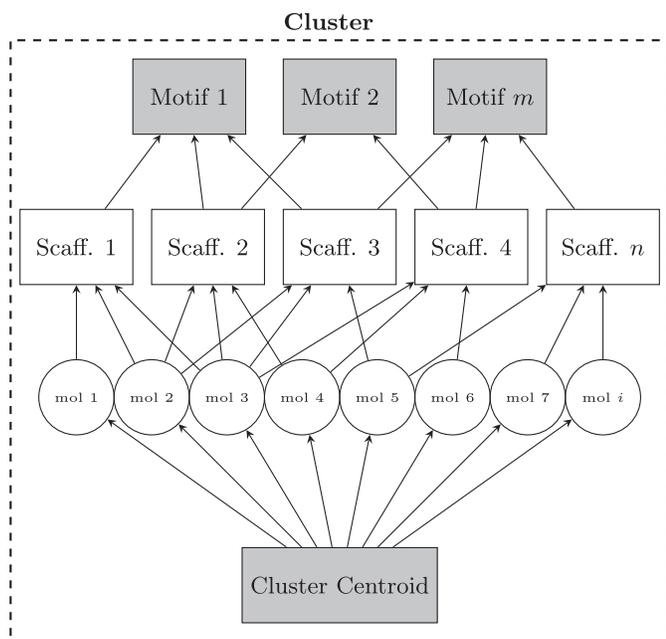

FIG. 10. Diagram of the motif generation and relationships to other molecular features.

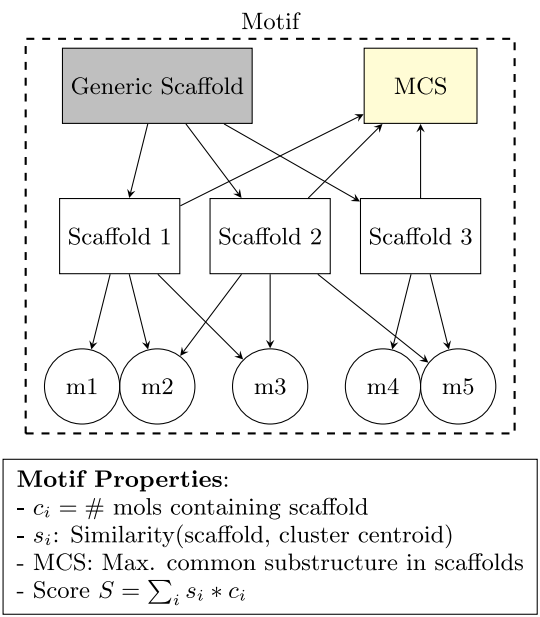

FIG. 11. Diagram to show scaffold and most common structure generation.

When forming motifs, we use objects known as scaffolds. Heuristically, scaffolds are graphs that represent collections of molecules within a defined similarity radius of a centroid, i.e., representing common molecular cores in a cluster. Unlike backbones or motifs, scaffolds are simple graphs with generic structures. Motifs group scaffolds by a single generic scaffold representing the maximum common substructure within the group.

### 2. Scaffold analysis

For each cluster, we select the 100 molecules closest to the cluster centroid by Tanimoto similarity. Note that the centroid is an abstract point in molecular fingerprint space, calculated to minimize the distance to all molecules in the cluster. Next, we generate Murcko scaffolds for these 100 molecules. Murcko scaffolds are a way of representing the core framework of a molecule—essentially the rings and linkers that remain after the removal of the side chains and the hydrogen substitute groups—providing a simplified "backbone" for structural comparison [29]. If the group contains multiple scaffolds, there can be significant degeneracy between them, e.g., two scaffolds differing by only a single atom or bond replacement.

To generalize beyond this degeneracy, we group scaffolds that are identical when converted to *generic* Murcko scaffolds. This process involves converting all atoms to carbon and all bonds to single bonds, i.e., generating the underlying basic graph. Finally, we compute the maximum common substructure (MCS) for each scaffold group. This representative structure, which may or may not contain wildcard atoms and/or bonds, is what we will call a motif.

We keep track of the similarity between each scaffold and the cluster centroid $s_i$. Then, by the number of molecules in the cluster that contains a given scaffold in the group $c_i$, we compute a group score as: $S = \sum_i s_i \cdot c_i$. In Fig. 11 we demonstrate the full scaffold grouping algorithm.

### 3. Cross-cluster analysis and molecular motifs

After analyzing individual clusters, we continue to a global analysis fully detailed in Fig. 12. Unsurprisingly, the same motif can exist within more than one cluster, though the cluster-specific scores are generally different. We would like to learn which motifs contribute globally to the clusters of candidate molecules. Therefore, we collect and merge motifs across all clusters. This process proceeds as follows:

(1) Collect scaffold groups from all clusters
(2) Merge groups with identical generic scaffolds
(3) Recalculate properties for merged groups
(4) Analyze molecular properties within each group

In merging motifs from different clusters, it is important to keep track of the cluster-specific properties. Therefore, the new score is computed based on the similarity between the underlying scaffolds and the centroid of their originating centroid. This conserves the relative importance of each motif to each cluster. As before, a new MCS is computed for the new motif. Finally, we can compute the average transition energies and oscillator strengths for each motif by averaging over the molecules that contain their underlying scaffolds.

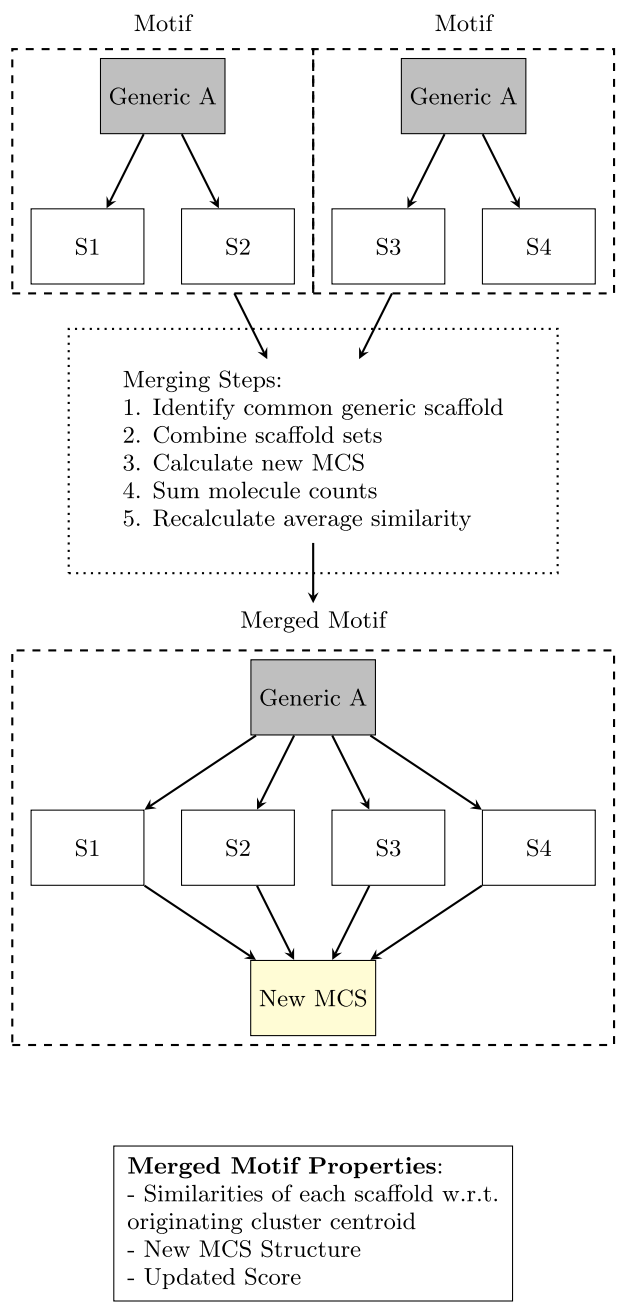

FIG. 12. Motif acceptance diagram.

## B. Crystal-driven analysis methodology

Using the molecular motifs, we are able to probe another desirable feature of a DM detector material, the ability to grow crystals. We have not trained our networks to optimize for this property, however, taking a dataset of known organic molecular crystals, we can attempt to glean which of these crystals likely fit our criteria by searching for our motifs' presence within these crystals. The CSD_EES_DB dataset, presented in [30], contains a list of known organic molecular crystals. We compute the previously mentioned group score of every element in the

CSD_EES_DB dataset to the top 300 most populated motifs and append the elements with the top group scores. Furthermore, we retain a list of more exotic crystallisable molecular structures from the CSD_EES_DB dataset after we remove motifs from our dataset which are contained within more than 100 crystals in Appendix D Fig. 31.

### C. Backbone-driven analysis methodology

The motif-driven analysis verifies key molecular structures related to our region of interest. However, to obtain specific molecular structures within the motifs, we must turn to the backbone-driven analysis. Like the motif-driven analysis, the backbone-driven analysis will have three main stages: clustering, backbone generation, and joint backbone and motif analysis.

#### 1. Clustering

What we reference as molecular backbones are common structures among the outputs of the SS algorithm, but which are more specific than the molecular motifs. Clusters associated with molecular backbones are of O(10–100), as opposed to the O(100–1,000) clusters generated by the `BitBIRCH` algorithm. They can be considered more specific examples from molecular classes that display the desired features.

We begin, like with the motif-driven analysis, by clustering our output from the SS algorithm. However, unlike the motif-driven analysis, we perform no cuts to the SS outputs since we are interested in very specific structures, regardless of synthesizability. A relatively high Tanimoto similarity threshold of 0.7 is set such that we do not risk the chaining effect, i.e., where two dissimilar clusters are joined due to two constituent molecules being somewhat similar. An example of how linked molecules are formed into clusters is shown in Fig. 13.

#### 2. Backbone generation

After forming the clusters, we proceed to identify the characteristic backbones which represent our clusters. We have established two categories of molecular backbones: simple and complex.

In Fig. 14 we show our generation and selection process. On the one hand, complex backbones are derived from `RDKit`'s `FindMCS` function. Molecules of a cluster are fed into this function to give the maximum common structure among the inputted molecules. This works well, however, misses out on simple structures that would be better suited to describe the cluster. Simple backbones are generated, on the other hand, by finding the Murcko scaffolds of molecules within a cluster and then finding the maximum common structure among the Murcko scaffolds using `RDKit`'s `FindMCS` function. Whichever backbone has the higher mean Tanimoto similarity to the cluster is chosen as the cluster's backbone. Clusters with the same backbones are then merged to form our final backbone lists.

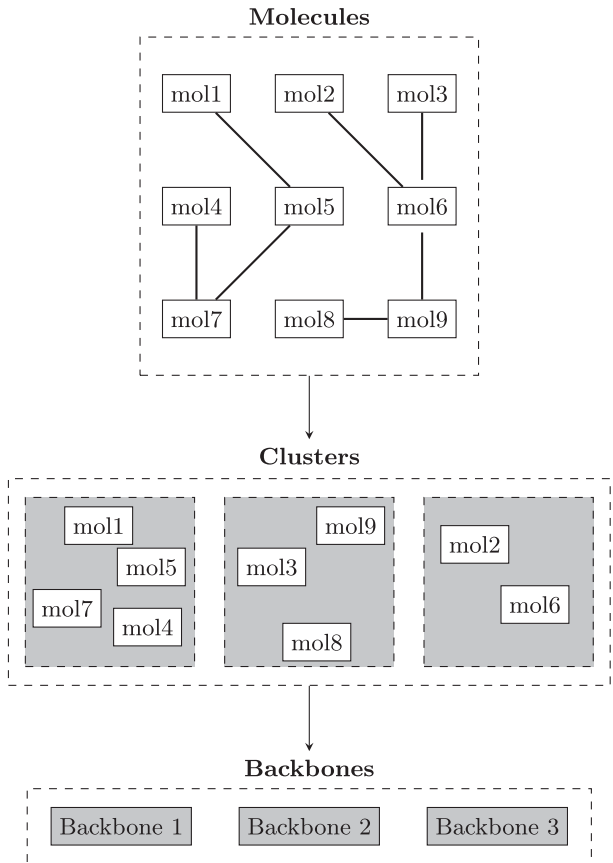

FIG. 13.   Diagram of the `Single-linkage` clustering algorithm and backbone generation.

## VII. MACHINE LEARNING AND SEEDED SAMPLING RESULTS

In this section, we will demonstrate the performance of our methods/network architectures and present findings from our seeded sampling algorithm on the PubChemQC3M dataset. To compare with the literature standards, we present the benchmark results for our MLP performance on the QM9 dataset in Appendix B, explicitly we show the performances via Figs. 24 and 25.

### A. Regression performance

In this part of the section, we will show the performance of the MLP on predicting properties for the lowest two excited states, the first and second transition energies ($\Delta E_{1,2}$) and the first and second oscillator strengths ($OS_{1,2}$).

In Table I, we show the performance of the MLP by displaying the training $R^2$ and the validation $R^2$ on the specific molecular properties. We give further validation plots in Figs. 26–29.

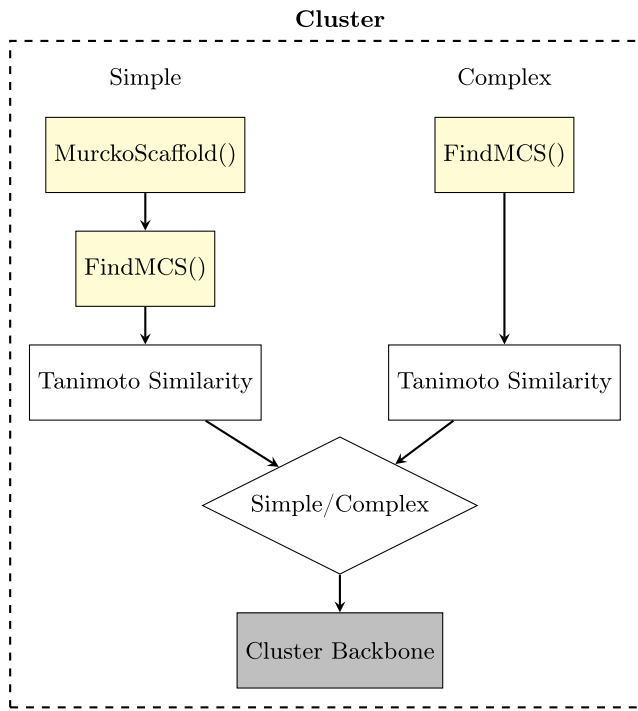

FIG. 14.   Diagram of the backbone generation process, showing the two classes of backbones: complex and simple.

Most importantly, we will show the MLP's ability to classify molecules as having desirable molecular properties. We defined desirable molecular properties as whether a molecule is predicted to meet the following criteria:

$$(\Delta E1 < 2.5 \text{ \& } OS1 > 0.05) \text{ \& } /$$
$$\text{or } (\Delta E2 < 2.5 \text{ \& } OS2 > 0.05)$$

In Fig. 15 we show our capacity to classify molecules as having desirable molecular properties. From this confusion matrix, we conclude that roughly half the molecules generated by the seeded sampling algorithm will have desirable molecular properties. A powerful feature is that the rate of false positive predictions is extremely low, well below the one percent mark. In other words, we are able to accurately throw away the overwhelming majority of molecular space, that constitutes the hay stack, while keeping almost the entire needle. Further confusion matrices demonstrating the performance of classifying individual molecular properties are shown in Appendix C 3 Fig. 30.

TABLE I.   Training and Validation $R^2$ values for different Networks on the PubChemQC3M dataset.

| Molecular property | Training $R^2$ | Validation $R^2$ |
| --- | --- | --- |
| $\Delta E1$ | 0.9788 | 0.9438 |
| $\Delta E2$ | 0.9765 | 0.9392 |
| OS1 | 0.7880 | 0.7478 |
| OS2 | 0.7603 | 0.5619 |

**Ideal molecules confusion matrix**

FIG. 15. Confusion matrix of classification for molecules with desirable molecular properties.

(i) Recall: 26.79%
(ii) False positive rate: 0.056%
(iii) Precision: 49.93%

### B. Seeded sampling performance

The seeded sampling algorithm has produced 8,400,160 molecules with predicted desirable molecular properties. This entire dataset is available at [6] under `ideal_mols_2`.

Our clustering algorithms produced two classes of molecular structures: *Motifs* and *Backbones*. Through the `BitBIRCH` clustering algorithm, 32,900 motifs were found. By nature, all SS outputs were clustered to at least one of the motifs. The `Single-linkage` clustering algorithm generated 125,628 molecular backbones. However, only 780,913 SS outputs were clustered, i.e., 9.30% of the generated dataset. The rest of the molecules were unique enough to this clustering, that no major common patterns were identified.

In addition, the molecular backbones (found using `Single-linkage clustering`) and their constituents can be found under `Cluster output folder` at [6].

### C. Benchmarks

The task of predicting molecular properties has attracted significant attention in the recent past. Numerous studies have tested various NN architectures to perform this task, and we do not present a comprehensive overview of those. A very recent example, however, is a study that found that tuning an existing large-language model (LLM), such as GPT-3 can produce a model that can successfully predict molecular properties, such as the HOMO and LUMO energy levels [31]. In this study, the SMILES language has been used to encode the molecular structures as an input for the LLM. However, we find that the reconstruction scores did not reach as high values as the networks presented in this study. In Ref. [32] a number of ML techniques, including random forest, and correlation-based feature selection algorithms, are used to learn chemical properties from the QM9 dataset. The work provides an overview of the mean average errors (MAEs) in the supporting material of Ref. [32] in Table III. The considered methods lead to MAEs of the order of 0.09 eV for ground-state energies, and ∼0.6 Debye for the lowest oscillator strength (Table II).

More recently, Ref. [33] demonstrated that a dual-branched message-passing neural network trained on molecular properties containing 3-dimensional information, such as bond angles between the atoms, can improve the predictive power. Here MAEs of the order of ∼0.02–0.04 eV where reached for the ground state, and first excited energy levels on the QM9 data set. The ground-state dipoles had MAEs of the order of ∼0.03 Debye.

We validated the MLP trained on the latent space vectors and fingerprints that is used in this work on the QM9 data set. We find an MAE of ∼0.01 eV for the HOMO-LUMO energy gap, and an MAE of ∼0.5 Debye for the corresponding oscillator strength. It is not surprising that this MLP outperforms the models used in Ref. [32] that are trained on SMILES only. It has a slightly better performance on the energies than the network of Ref. [33], but is somewhat weaker on the dipole magnitude. That last fact is also expected, as three-dimensional information seems to be very beneficial for the reconstruction of this ground-state property.

### VIII. RESULTS

In this section, we present our final results. They assume three forms: motifs, crystals, and backbones. These three structures and their generation have been described in Sec. VI. However, for modularity, we will describe them again. Motifs are intended to be very general structures associated with low transition energies and high oscillator strengths. The crystal list we find is meant to be a list of known crystalizable molecules most strongly associated with our motifs, and therefore, our ROI. And the backbones are intended to be specific exemplary molecular structures that we find to have the molecular properties in our ROI. One can see further descriptions in Fig. 16.

We present three primary shortlists detailed by Fig. 16 in Figs. 17, 19, 21. The lists presented in Figs. 17 and 21 were found by ranking molecular motifs and backbones by the mean molecular properties within their clusters. For both the 300 most populated motifs and backbones with cluster sizes larger than 20, we generate four ordered lists corresponding to the four molecular properties of interest, i.e., 2 sorted lists where the oscillator strengths go top to bottom and 2 sorted lists where the transition energies go bottom to top.

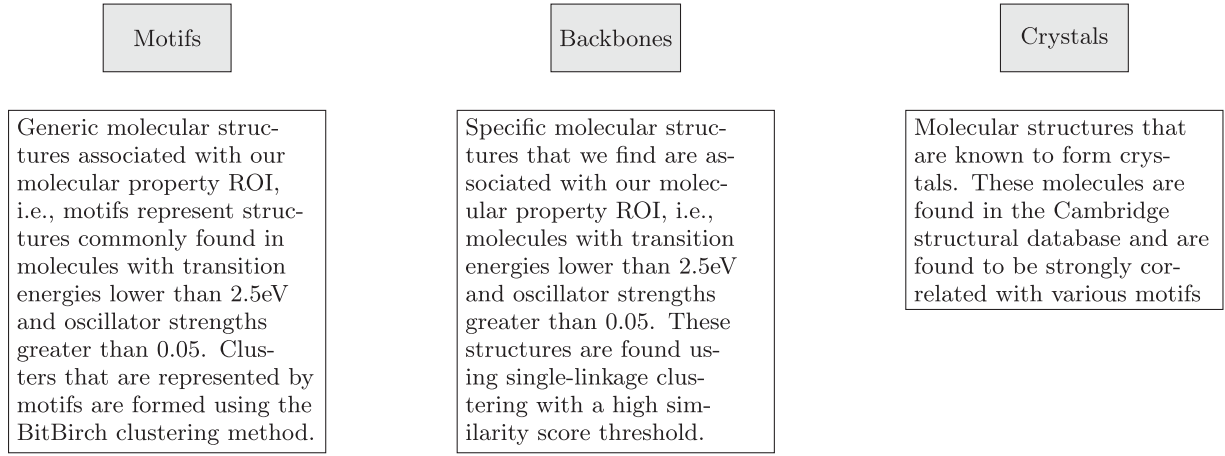

FIG. 16. This diagram shows the three primary results we generate from this paper: motifs, crystals, and backbones. Motif generation and crystal matching is discussed in Sec. VI A and backbone generation is discussed in Sec. VI C.

More explicitly, we begin by examining the top 10 motifs/backbones in each list. At each iteration, we increase the number considered from each list by one, continuing until 10 motifs/backbones are common to all four lists. These shared motifs/backbones represent those with the most ideal molecular properties on average.

### A. Motif results

Our primary molecular motif shortlist is shown in Fig. 17. These motifs represent general structures most strongly associated with low transition energies and high oscillator strengths.

The molecular motifs found in the cross-cluster analysis are sorted by descending score and the top 20 are shown in Fig. 18. As would be expected, single aromatic rings are ubiquitous. Previous experiments have found that even xylene, whose molecular structure is a benzene ring with two substituted methyl ($CH_3$) groups, is an organic scintillator with an intrinsic sensitivity to dark matter that is comparable to silicon-based detectors [4]. We find that polyacenes are also well-represented among molecular candidates, e.g., naphthalene (fits motif 3) and anthracene (fits motif 5). This agrees with the conventional wisdom of organic scintillating materials, where for example, naphthalene is used as a wavelength shifter in binary scintillators such as EJ-301. Similarly, we also find motifs in which ringed groups are bridged by a molecular chain hosting delocalized electrons. Previous work on, e.g., trans-stilbene (fits motif 13) has found that its anisotropic structure is particularly useful in dark matter searches [20]. However, as noted by previous studies, common organic scintillators such as xylene and trans-stilbene suffer from suppressed (and indeed forbidden) first transitions. Motif 13 suggests that replacing one of the carbons, in the central bridge, for a nitrogen, is a molecular "direction" that is also common with low-threshold transitions of high-probability.

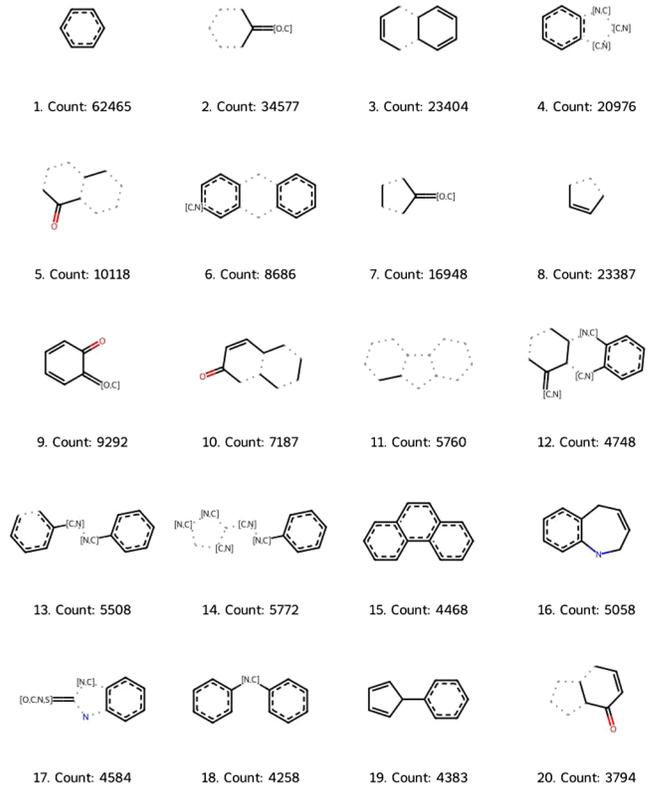

FIG. 18. Top 20 molecular motifs ranked by score. Note that certain atoms can be substituted by [C,N,O,S] as long as the structure remains isoelectronic. Dotted lines indicate that the bond could be a double or single bond. Count here refers to the number of molecules generated by the seeded sampling algorithm which have been identified as being associated with the specific structure.

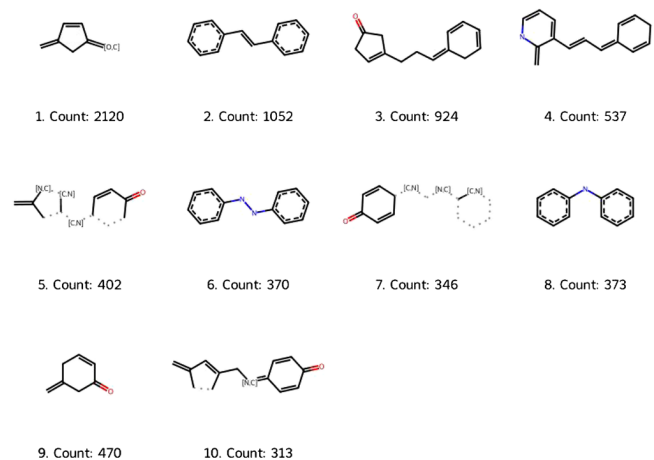

FIG. 17. Top 10 motifs ranked by the process described in the opening of Sec. VIII.

In general, we find that nitrogen can be substituted for carbon in certain rings in order to alter an aromatic polycyclic structure without destroying the delocalized system. Similarly, we find that oxidizing carbons in phenyl groups (motifs 2, 5, 7, 9, 10, etc.), is also a desirable substitution. We suggest this is due to the distortion of the electron density away from symmetric rings. Immediately, we can suggest that perhaps dione-substituted stilbene and/or azo-stilbene derivatives may be an interesting class of molecules to study. The same combinatorial exercise can be carried out with many of these motifs in order to find classes of molecules that would otherwise not be considered as dark matter-detector targets, e.g., phenanthrene quinone. One of the key takeaways from this analysis is that planar polycyclic molecules with extended delocalized pi-electron systems are generally expected to fit our criteria, provided they do not have symmetry-forbidden transitions. For this reason, we suggest that asymmetric motifs are of particular interest such as motifs 11, 14, and 15. One could also consider symmetric motifs and construct derivatives that further extend the pi-electron system in an asymmetric way, as with the relationship between motifs 3 and 10.

## B. Crystal results

Following from the primary molecular motif list, we generate a primary list of 10 organic molecular crystals from the CSD_EES_DB dataset in Fig. 19 which are most strongly associated with our molecular motifs. As such, we identify these crystal structures as being associated with low transition energies and high oscillator strengths.

Of the molecules present in Fig. 19, the common pattern is polyphenol rings, with large delocalized pi-electron systems. For all but two exceptions, all identified motifs have a structural asymmetry or contain an atom with large electronegativity that draws the electron cloud and creates an asymmetric electron density distribution. Overall, it is very pleasing that among the suggested molecules with the desired properties, a substantial number of candidates are known to form crystals. We plan to further investigate

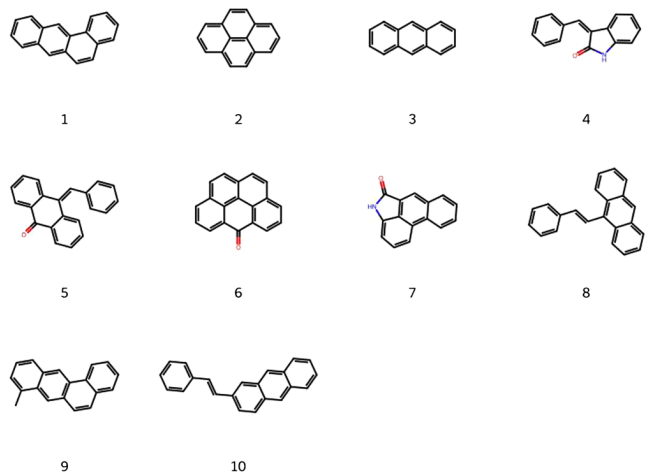

FIG. 19. Top 10 organic molecular crystals from the CSD_EES_DB dataset, ranked by group score and compared to the 300 most populated motifs.

whether this property can also be optimized in a forthcoming study.

## C. Top backbones

Like with the motifs and crystals, we present 10 backbone structures most heavily associated with low transition energies and high oscillator strengths as shown in Fig. 20. Unlike the motifs, these structures are very specific and are meant to be interpreted as specific structures one would ask a chemist to synthesize (with minor modifications).

From the figures, Figs. 17 and 21, it is easy to see how motifs describe general features within our region of interest and backbones describe specific, strong features. The utility of the backbones becomes clear when analyzing the constituent molecules, as shown in Fig. 32. When analyzing ranked motif 6, one would have to trudge through 370 molecules to find an optimized derivative. However, ranked backbones 8 and 9 are explicitly

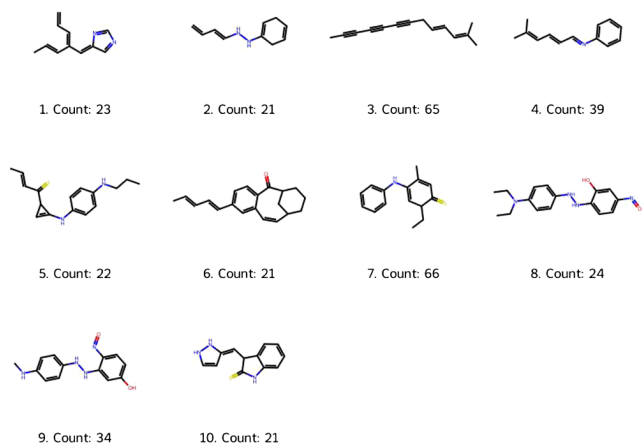

FIG. 20. Top 10 backbones ranked by the process described in the opening of Sec. VIII.

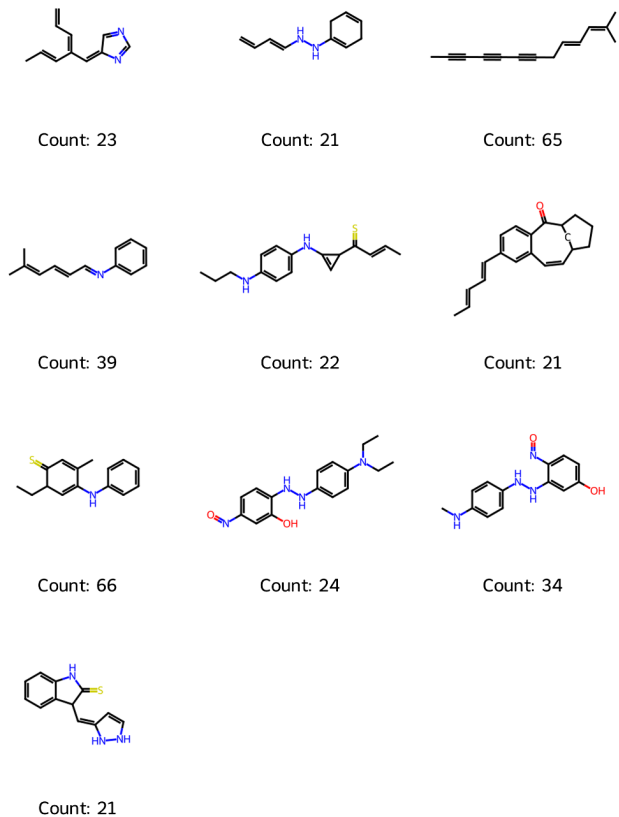

Count: 23

Count: 21

Count: 65

Count: 39

Count: 22

Count: 21

Count: 66

Count: 24

Count: 34

Count: 21

FIG. 21. Top 10 backbones ranked by the process described in the opening of Sec. VIII.

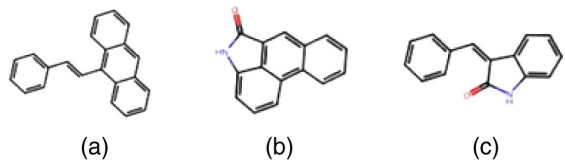

(a)            (b)            (c)

FIG. 22. Three molecular structures that are known to form crystals, and feature typical patterns that our network identified as promising. Large delocalized $\pi$-electron systems, and asymmetries in the molecular backbone. Those crystals are available and will undergo laboratory testing shortly.

optimized derivatives of ranked motif 6. It is likely that sampling within any one of these molecular backbones would ultimately lead to the same results and so the top molecular backbone list serves as an extremely useful pointer to molecular structures that are likely to be optimal.

## IX. DISCUSSION

In this proof-of-principle work, we have demonstrated that a combination of generative and regressive AI models, a VAE and MLP, can be used to predict novel molecular structures from which we are able to select candidates with optimal proxy properties for dark matter direct detection.

At the core of the method lies the molecular language of SELFIES which has an advantageous property of being robust under point mutations, which greatly adds to the stability of the generation process. In addition, the high level of abstraction allows an efficient one-hot embedding, and again an efficient way of generating the VAE latent space.

However, as we have discovered, latent space vectors roughly match the information density of their inputs and, as such, do not provide sufficient performance for molecular property predictions. Luckily, a workaround exists. As in the case of natural language, a sentence—that is a structure that is easy to generate based on semantic rules—can be translated into an object with higher information density, such as a descriptive image. Similarly, in our approach, the latent space vectors are decoded back into the molecular string language, and then converted by several known means into fingerprint representation of the molecules. We find that using the fingerprint representations, rather than the latent representation as inputs, significantly boosts MLP performance, particularly for the oscillator strengths and second transition energies. An explicit comparison between two different MLPs are shown in Appendix F Fig. III, where the MLPs are trained on either only the encoded $\mu$s of the molecules alone or with the addition of fingerprints and mol2vec representations.

After identifying this ideal architecture we were able to run it using a seeded sampling algorithm that generated millions of candidate molecules that are expected to have the desired properties. In order to process this vast output, we have implemented clustering techniques, that identified a number of molecular backbones and motifs, that constitute structures that are expected to be common in ideal molecules for dark matter detection. As discussed in Sec. VI A 3 a common feature of optimal molecular structures, is a large delocalized system of pi-electrons, and an asymmetry in the electron density distribution, a feature that likely prevents low-lying transitions from being classically forbidden, i.e., identically zero oscillator strength.

In Fig. 22, we show three promising structures that show the above features, and are also known to form crystals. As those molecules are available in crystalline form the next step will be a first experimental validation of their properties, that are predicted by our ML architecture. In future works, we will develop this validation step within the pipeline to check that our machine learning predictions match the TDDFT computed predictions.

## X. CONCLUSION

In this work, we have successfully demonstrated how a generative neural network, working together with a network trained for property prediction can generate lists of molecular candidates with desired properties. We are interested in identifying optimal molecular targets for sub-GeV DM detection, similar to the strategy outlined in Ref. [7]. Thus, we have focused our attention on molecules with low excitation energies, and large oscillator strength, which are good estimators for molecular matrix elements relevant for dark matter detection. We employed a sampling algorithm that allowed us to generate a large number of potential candidate molecules and employed a couple of clustering prescriptions to identify the common backbones and motifs of the relevant molecular structures.

Ultimately, we present three shortlists of molecules / molecular structures in Figs. 17, 19, and 21 which are potential ideal organic detector targets for sub-GeV dark matter detection. In upcoming work, we will employ methods of quantum chemistry and chem-informatics to validate our predictions in order to identify candidates for laboratory testing. We found that a few of our candidate molecules are known to form crystals, which is a promising sign indicating that we are coming closer to our goal of identifying optimal materials for new, inexpensive dark matter detectors.

## ACKNOWLEDGMENTS

We thank Louis Hamaide, Stefan Nietz and Samuel Godwood for helpful comments on the draft. We thank Samuel D. McDermott for very useful advice and help with our codebase. The authors are grateful to the Mainz Institute for Theoretical Physics (MITP) of the DFG Cluster of Excellence PRISMA+ (Project ID 39083149), for its hospitality and its partial support during the completion of this work. The work of C. B. was supported in part by NASA through the NASA Hubble Fellowship Program grant HST-HF2-51451.001-A awarded by the Space Telescope Science Institute, which is operated by the Association of Universities for Research in Astronomy, Inc., for NASA, under contract No. NAS5-26555 as well as by the European Research Council under Grant No. 742104. We acknowledge support from J. S.'s UK Research and Innovation Future Leader Fellowship MR/Y018656/1.

## DATA AVAILABILITY

The data that support the findings of this article are openly available [6].

## APPENDIX A: NETWORK DIAGRAMS

This section shows the full seeded sampling flow diagram. The process can also be described fully as follows:

(1) We take our molecular seed list and encode them into $\mu$s and $\sigma$s, focusing on the means. Each are represented as tensors of size [N, D] where N is the length of the molecular seed list and D is the dimension of the latent vectors used.

(2) We take the $\boldsymbol{\mu_N}$, where N is the step in the loop. If we have just begun, then this means that we take the first N, i.e., the 1st element of our [N, D] tensor.

(3) Inside of our loop, we sample from the distribution described by the corresponding $\sigma$ tensor, forcing the space in which we explore to be

$[1\sigma < |\mathbf{z} - \mu| < 2\sigma]$. Nominally, we generated 100,000 of these $\mathbf{z}$s per given molecular seed. I.e., from 1 $\mu$ corresponding to a molecular seed, we generate 100,000 $\mathbf{z}$s.

(4) For a given $\mathbf{z}$, once pushed through the VAE decoder, we generate a probability vector of the size of the input one-hot vector, i.e., [X,Y] where X is the length of the longest SELFIES input and Y is the length of the SELFIES alphabet. The probability vector can be converted to a sequence vector that corresponds to SELFIES alphabet indices by taking the `argmax()` over the Y dimension. This sequence effectively corresponds to a SELFIES where, instead of SELFIES characters, we have indices corresponding to elements of the SELFIES alphabet.

(5) By taking the sequence tensor, we can easily remove degenerate sequences, thus removing many nonunique SELFIES/SMILES.

(6) It is then trivial to decode the SELFIES sequences to SMILES by generating SELFIES strings. Once we have obtained the SELFIES, we use the SELFIES package to generate corresponding SMILES.

(7) We sanitise the molecules by doing quick checks on the SMILES geometry, checking that decoded SELFIES characters are contained within the original SELFIES alphabet, canonicalizing the SMILES (which further removes redundancy). After this, we check that we still have molecules left.

(8) Taking the legitimate SMILES, we convert to $\mu$s, mol2vecs, Morgan fingerprints and Daylight fingerprints. Once we have these, we push them into the MLP to predict their molecular properties. It is important to note that MLP performance always decreased if we lowered the molecular fingerprint size and/or reduced the number of fingerprints used.

(9) If our molecules have desirable predicted molecular properties then we append them.

(10) If the loop has not yet been completed, we add 1 to the SMILES index. Once the loop finishes, we scan the entire dataset, removing degeneracy and calculating SA scores.

(i) Threshold 1: Number of remaining SMILES is larger than 0.

(ii) Threshold 2: $(\Delta E_1 < 2.5 \, \& \, OS_1 > 0.05) \, \& \, /OR \times (\Delta E_2 < 2.5 \, \& \, OS_2 > 0.05)$

(iii) Threshold 3: Number of remaining molecules in seed list = 0

## APPENDIX B: QM9 BENCHMARKS

Performance on the QM9 dataset is widely accepted as an important benchmark for machine learning models trained to predict molecular properties. Here, we show our predictive power on the HOMO-LUMO gaps and ground-state dipole moments of molecules within the QM9 dataset. The HOMO-LUMO gap and the ground-state dipole moments serve as useful analogs to the transition energies and oscillator strengths used in our main pipeline.

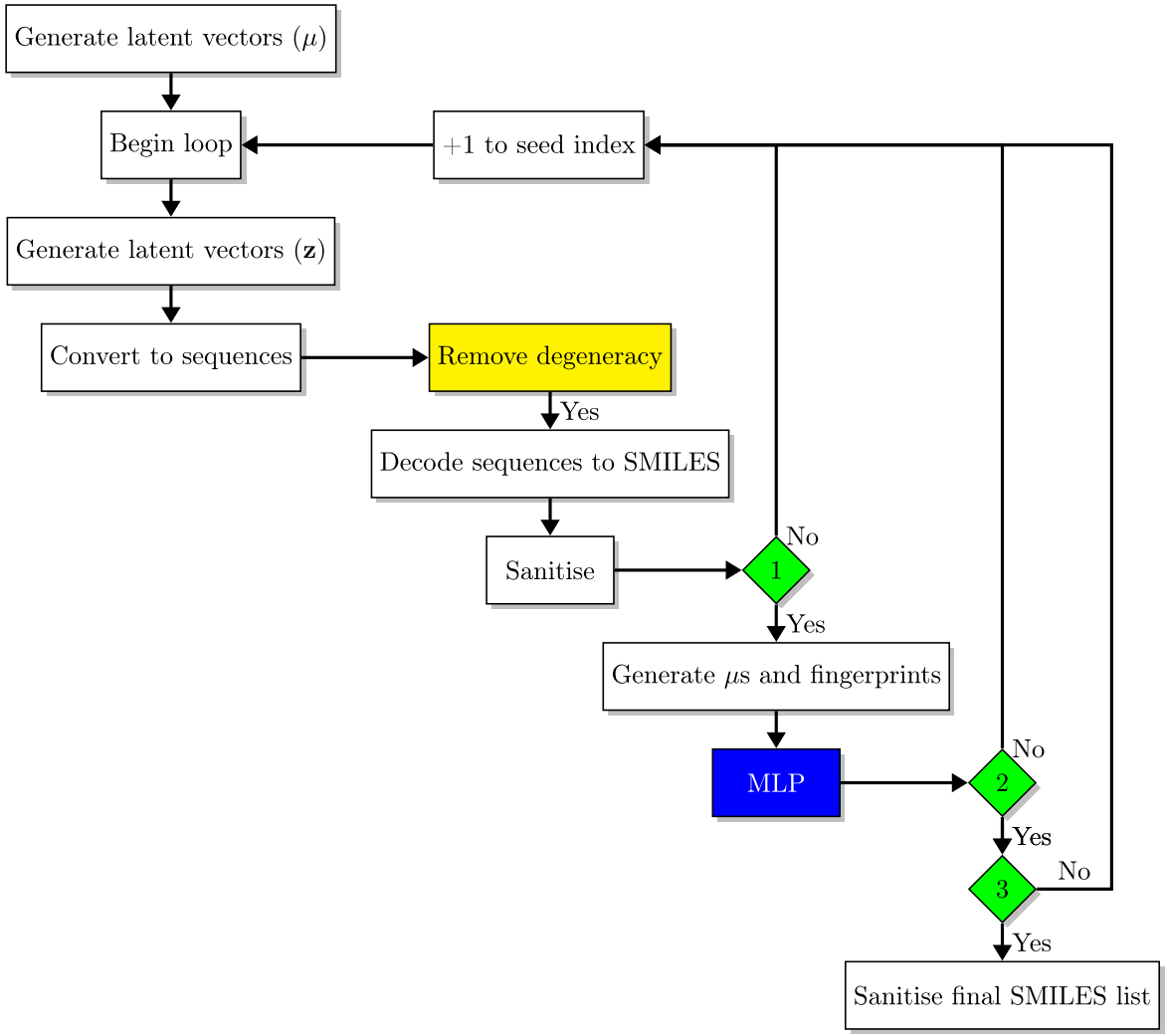

FIG. 23. Flow diagram of the seeded sampling process. The decision nodes reference whether there are any molecules that meet the threshold. Thresholds are given below, e.g., the first threshold checks that the number of molecules still around is greater than 0.

We show the training and validation $R^2$s and show plots to demonstrate our performance in various ranges. Notably,

TABLE II. Training and Validation $R^2$ values for the HOMO-LUMO gap and ground-state dipole moment properties by a model trained on QM9.

| Property | Training $R^2$ | Validation $R^2$ |
|---|---|---|
| HOMO-LUMO gap | 0.9452 | 0.9362 |
| Dipole moment | 0.9694 | 0.7467 |

our training $R^2$ performance is fairly high for the ground-state dipole moments. This is not particularly important since, for our purposes, we only care about validation $R^2$, which is still fairly high. In general, we see that the ground-state dipole moment, like the oscillator strengths, are more difficult to train on than the HOMO-LUMO gap, the transition energy analog.

## APPENDIX C: MLP PERFORMANCE PLOTS

Here, we will show the specific performances for all of the MLPs used within this paper. All of these plots are generated by their respective validation datasets.

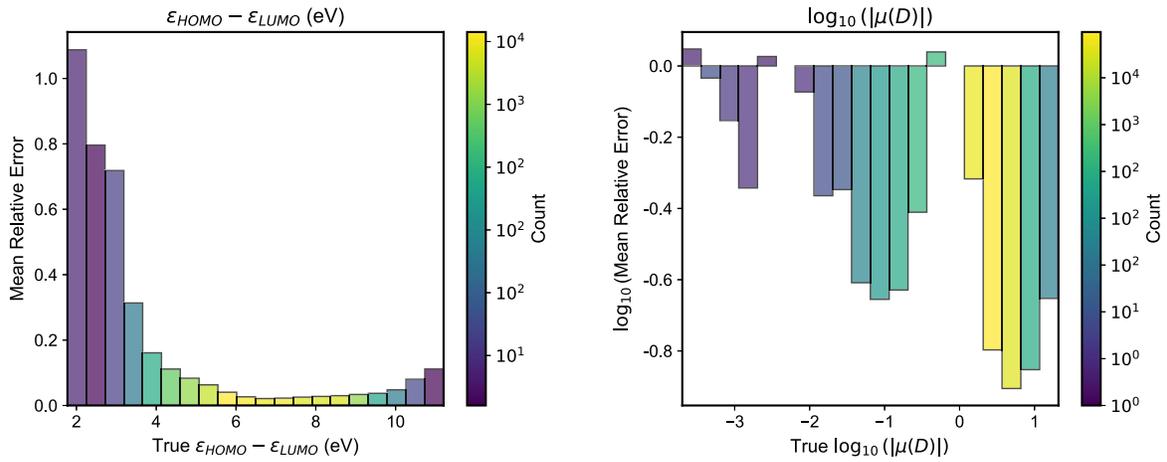

FIG. 24. We show the distribution of mean relative errors of the true values for the transition energies of our validation set of QM9. The plot on the left shows the HOMO/LUMO gap performance and the plot on the right shows the ground-state dipole moment performance.

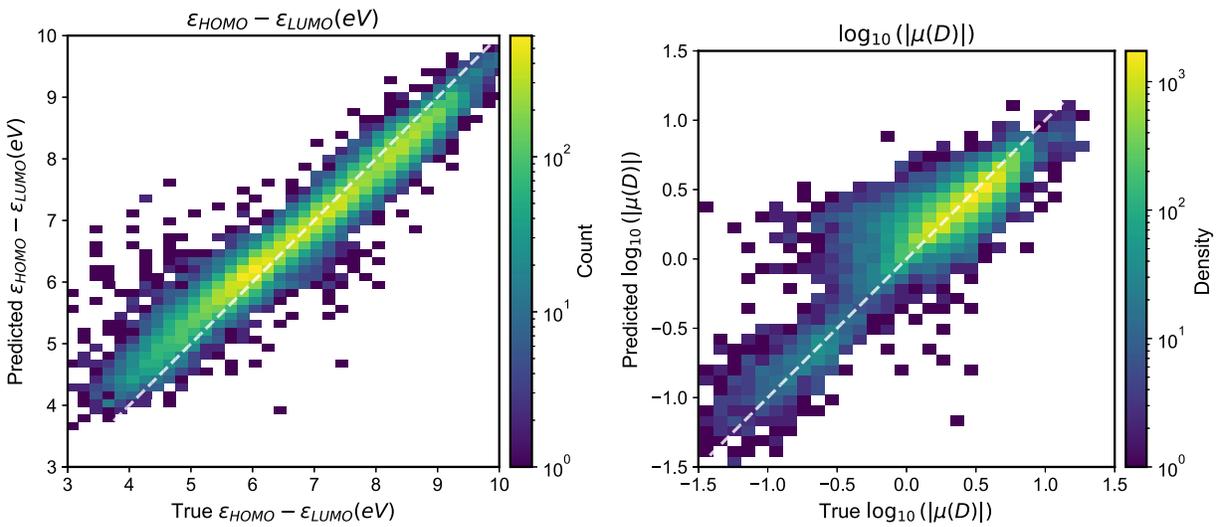

FIG. 25. We show the density of predicted values vs true values for the HOMO/LUMO gap and ground-state dipole moment in our validation set of QM9.

## 1. Bar charts

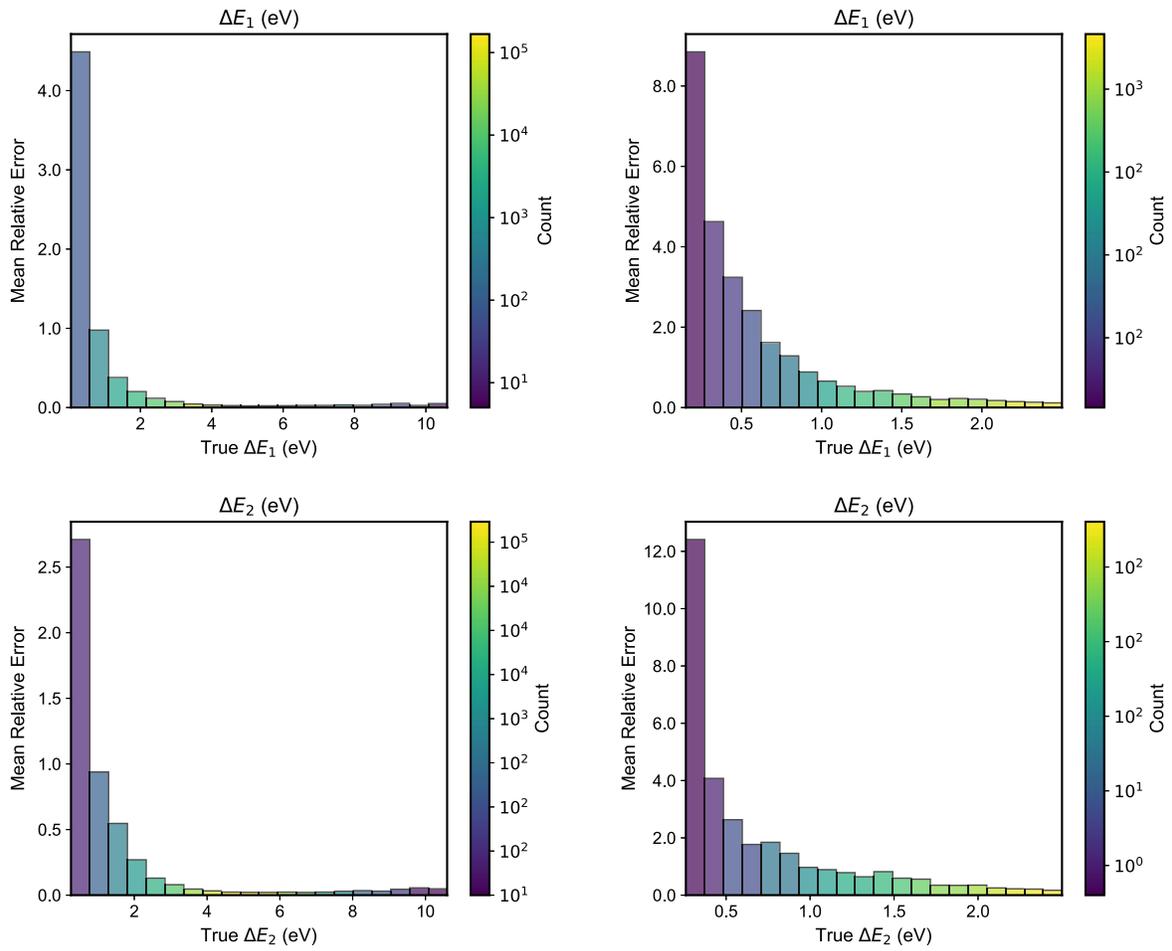

FIG. 26.   We show the distribution of mean relative errors of the true values for the transition energies of our validation set. Plots on the left show the distribution between the entire dataset and plots on the right only show the distribution of true values for molecules whose predicted value is below the threshold.

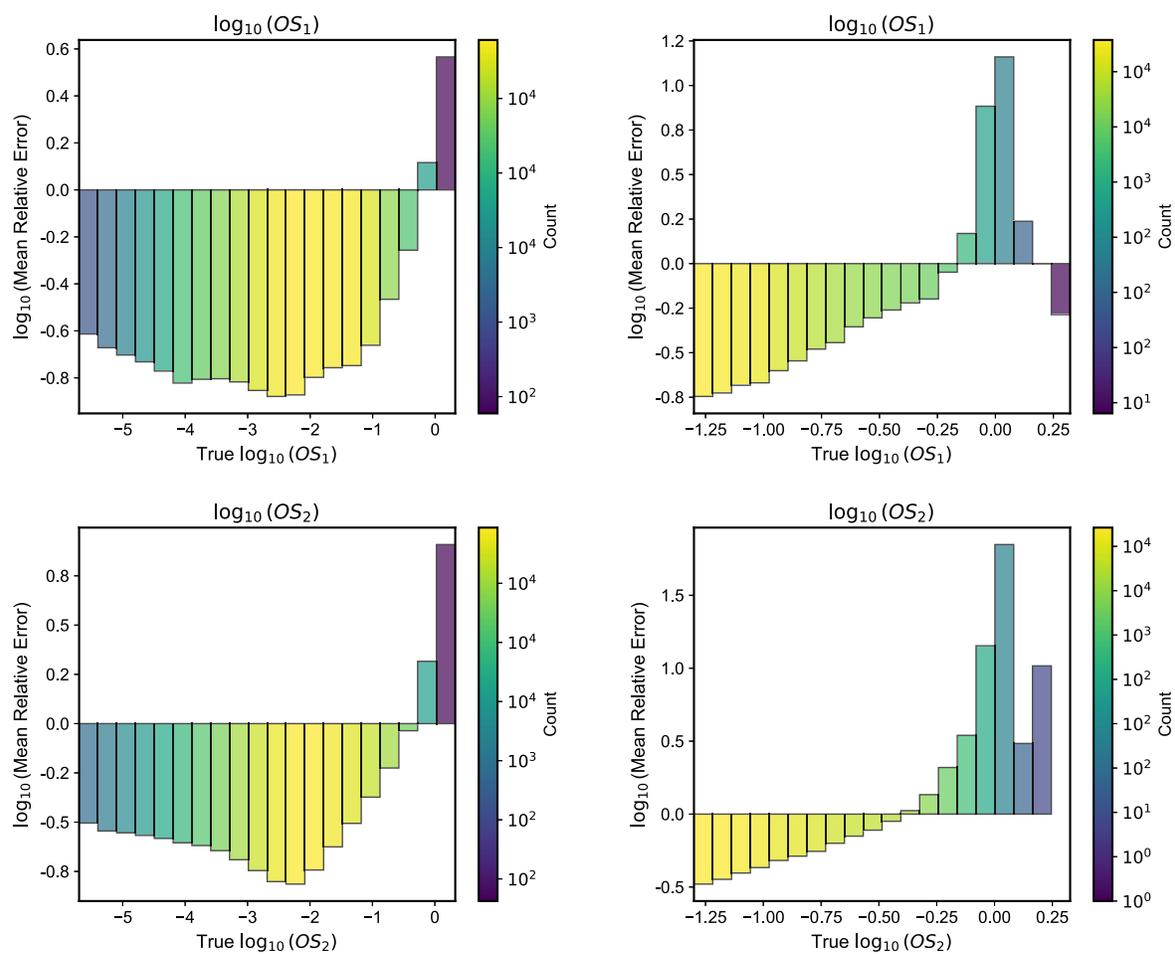

FIG. 27.   We show the distribution of mean relative errors of the true values for the oscillator strength of our validation set. Plots on the left the distribution between the entire dataset and plots on the right only show the distribution of true values for molecules whose predicted value is below the threshold.

## 2. Thresholds

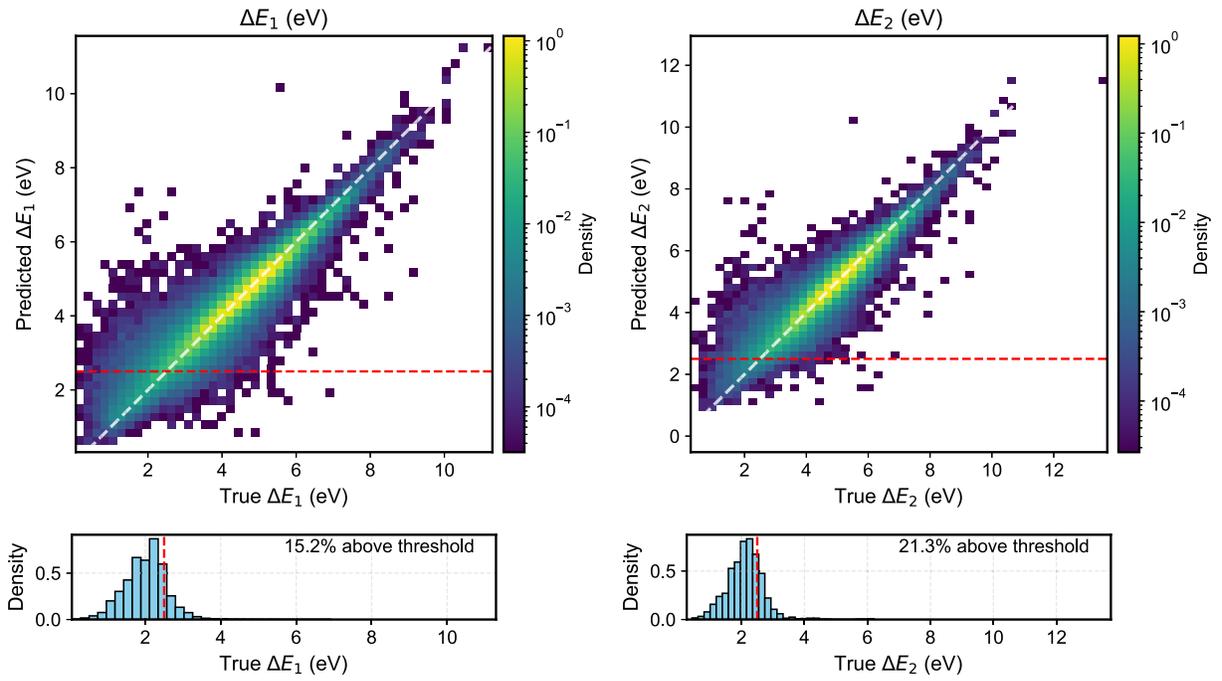

FIG. 28. We show the density of predicted values vs true values for the transition energies of our validation set, with a red line that corresponds to the threshold placed on the predicted value. The histogram shows the distribution of true values for molecules whose predicted value is above the threshold.

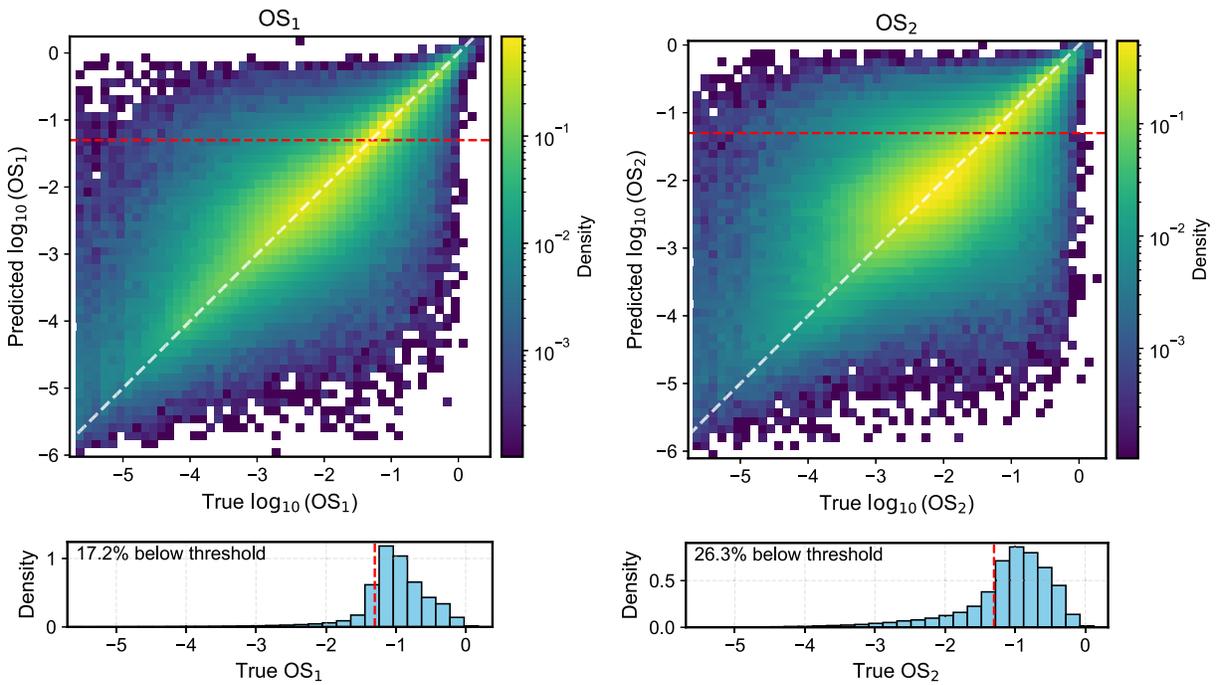

FIG. 29. We show the density of predicted values vs true values for the oscillator strength of our validation set, with a red line that corresponds to the threshold placed on the predicted value. The Histogram shows the distribution of true values for molecules whose predicted value is above the threshold.

### 3. Confusion matrices

In Fig. 30 we show how our predictions differ from the truth values which meet our thresholds. The matrices contain four fields with the following entries. Top left quarter (1, 1): This shows molecules with predictions that meet our threshold and also have truth values that meet our threshold. Top right quarter (0, 1): This shows molecules with predictions that meet our threshold but have truth values that do not meet our threshold. Bottom left quarter (1, 0): This shows molecules with predictions that do not meet our threshold but have truth values that do meet our threshold. Bottom right quarter (0, 0): This shows molecules with predictions that do not meet our threshold and also have truth values that do not meet our threshold. Summarizing the results of the confusion matrices we find the following performance benchmarks for the four observables we considered

(i) Recall: 67.13% ($\Delta E_1$), 44.65% ($\Delta E_2$), 78.26% (OS1), 62.03% (OS2)

(ii) False positive rate: 0.25% ($\Delta E_1$), 0.07% ($\Delta E_2$), 4.22% (OS1), 5.42% (OS2)

(iii) Precision: 84.80% ($\Delta E_1$), 78.66% ($\Delta E_2$), 82.83% (OS1), 73.69% (OS2)

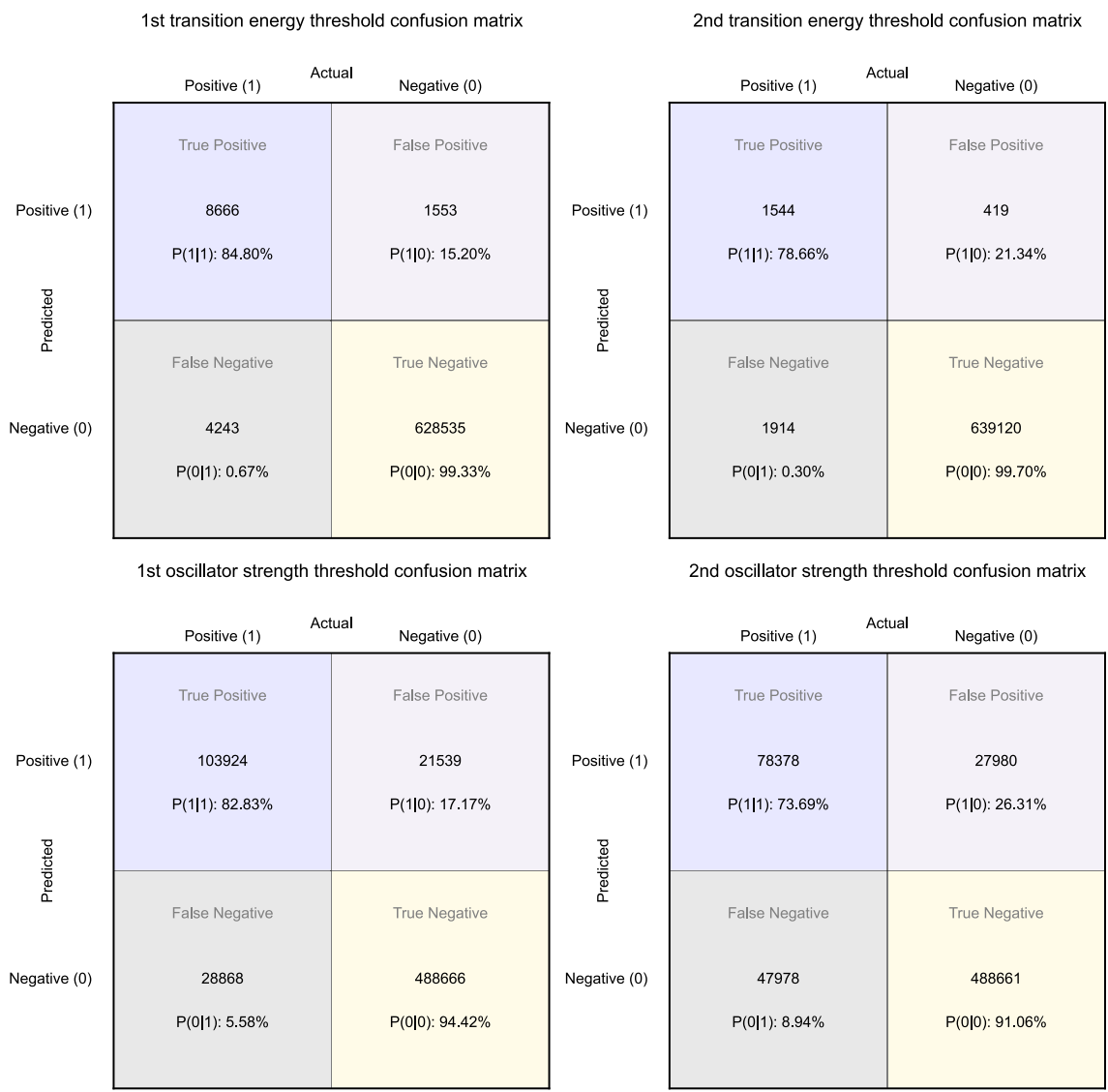

FIG. 30. These plots summarise our ability to classify individual molecular properties using our MLP as being below/above a given threshold. $\Delta E$ thresholds are below 2.5 eV and OS thresholds are above 0.05.

## APPENDIX D: EXTRA CRYSTALS

Here we show the extra, more exotic crystals found by removing common motifs from the group scoring system, i.e., we do not include scores associated with motifs with more than 100 matches to the crystal dataset.

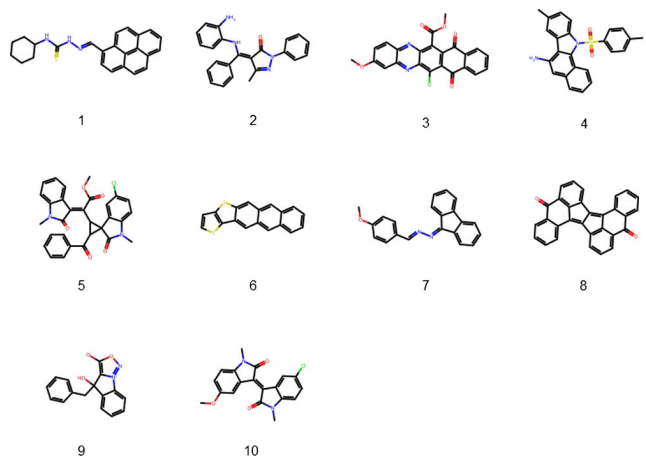

FIG. 31. More exotic crystals associated with the less popular general motifs.

## APPENDIX E: BACKBONE CONSTITUENT MOLECULES

Here, we show the constituent molecules which form the backbone 9 shown in Fig. 21.

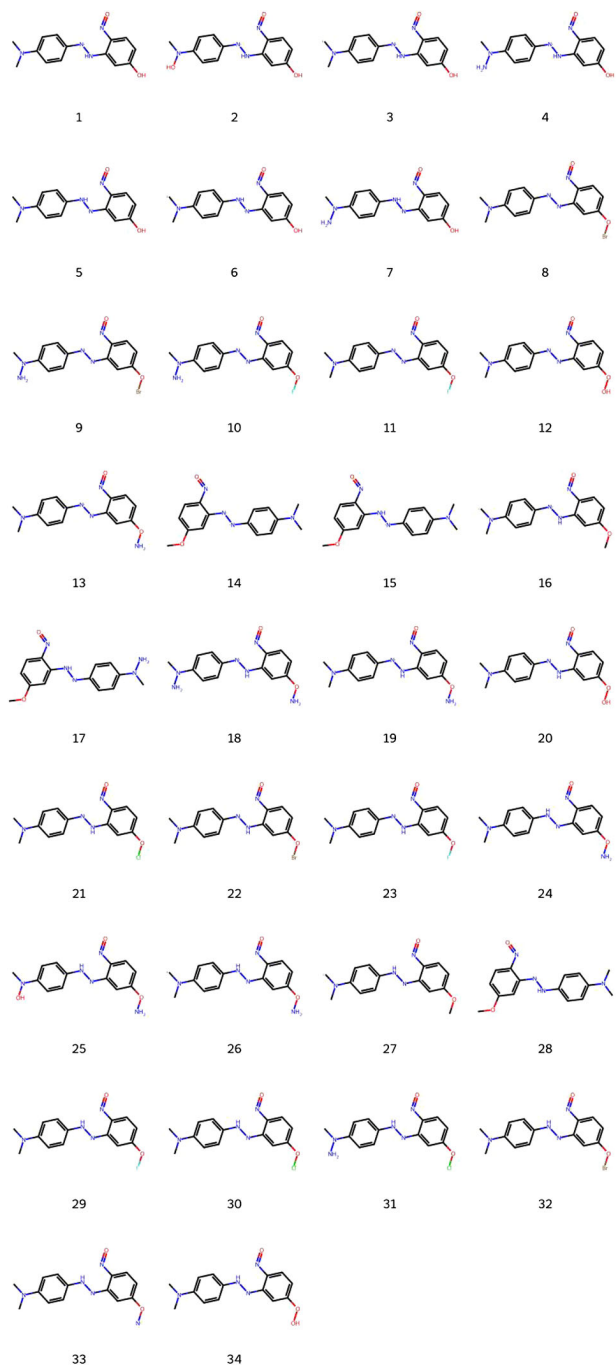

FIG. 32. SS outputs associated with the cluster which the 9th ranked backbone represents.

## APPENDIX F: FINGERPRINT VALIDATION

Here we demonstrate the utility of the fingerprints by comparing two MLPs with optimized hyperparameters with one trained on just the $\mu$ output from the encoder and one trained on the $\mu$, the fingerprints and the mol2vec embeddings. Predictions based on only the $\mu$ for the first oscillator strength were found to be very poor and so we do not show performances for molecular properties beyond the first transition energy here.

It is important to note that the hyperparameter search was not as extensive on the $\mu$ + FP + Mol2vec network due to the increased time for training. Furthermore, the network was optimized for all molecular properties,

TABLE III. Training and Validation $R^2$ values for predictions of the first transition energies of molecules in the PubChemQC3M dataset. The first entry shows performance of an MLP with optimized hyperparameters trained on only the $\mu$. The second entry shows the performance of an MLP with optimized hyperparameters trained on $\mu$ and molecular fingerprints.

| Network & Property | Training $R^2$ | Validation $R^2$ |
|---|---|---|
| $\mu$ only ($\Delta E_1$) | 0.8851 | 0.8292 |
| $\mu$ + FP + Mol2vec ($\Delta E_1$) | 0.9788 | 0.9438 |

not just the first transition energy, unlike the $\mu$ only network.